\documentclass[usenatbib]{mnras}

\usepackage{newtxtext,newtxmath}
\usepackage[T1]{fontenc}

\DeclareRobustCommand{\VAN}[3]{#2}
\let\VANthebibliography\thebibliography
\def\thebibliography{\DeclareRobustCommand{\VAN}[3]{##3}\VANthebibliography}

\usepackage{graphicx}	
\usepackage{amsmath}	

\usepackage{xspace,color}

\title[Chemical Abundance of the  LINER galaxy UGC\,4805.]{Chemical Abundance of the  LINER galaxy UGC\,4805 with SDSS-IV MaNGA}

\author[A.C. Krabbe et al.]{
A.C. Krabbe,$^{1}$\thanks{E-mail: angela.krabbe@gmail.com}
C. B. Oliveira Jr.,$^{1}$
I. A. Zinchenko,$^{2,3}$
J. A. Hern\'andez-Jim\'enez,$^{4}$ 
\newauthor{O. L. Dors Jr.,$^{1}$
G.~F. H\"agele,$^{5,6}$
 M.~V. Cardaci,$^{5,6}$
N. R. Telles$^{1}$}\\
$^{1}$ Universidade do Vale do Para\'{\i}ba, Av. Shishima Hifumi, 2911, Zip Code 12244-000, S\~ao Jos\'e dos Campos, SP, Brazil\\
$^2$ Faculty of Physics, Ludwig-Maximilians-Universit\"{a}t, Scheinerstr. 1, 81679 Munich, Germany \\
$^3$ Main Astronomical Observatory, National Academy of Sciences of Ukraine, 27 Akad. Zabolotnoho St 03680 Kyiv, Ukraine \\
$^{4}$ Departamento de Ciencias F\'{\i}sicas, Universidad Andr\'es Bello, Fern\'andez Concha, 700, Las Condes, Santiago, Chile.\\
$^{5}$ Instituto de Astrof\'isica de La Plata (CONICET La Plata--UNLP), Argentina. \\
$^{6}$ Facultad de Ciencias Astron\'omicas y Geof\'{\i}sicas, Universidad Nacional de La Plata, Paseo del Bosque s/n, 1900 La Plata, Argentina\\
}

\date{Accepted XXX. Received YYY; in original form ZZZ}

\pubyear{2020}

\begin{document}
\label{firstpage}
\pagerange{\pageref{firstpage}--\pageref{lastpage}}
\maketitle
\begin{abstract}
Chemical abundance determinations in Low-Ionization Nuclear Line Regions (LINERs) are especially complex and uncertain because the nature of the ionizing source of this kind of object is unknown.  In this work, we study the
oxygen abundance in relation to the hydrogen abundance (O/H) of the gas phase of the UGC\,4805 LINER nucleus. Optical spectroscopic data from the Mapping Nearby Galaxies (MaNGA) survey was employed to derive the O/H abundance of the UGC\,4805 nucleus based on the extrapolation of the disk abundance gradient, on calibrations between  O/H abundance and strong emission-lines for Active Galactic Nuclei (AGNs) as well as on photoionization models built with the {\sc Cloudy} code, assuming gas accretion into a black hole (AGN) and post-Asymptotic Giant Branch (p-AGB) stars with different effective temperatures. We found that abundance gradient extrapolations,  AGN calibrations, AGN and p-AGB photoionization models  produce similar O/H values for the UGC\,4805 nucleus and similar ionization parameter values.
 The study demonstrated that the methods used to estimate the O/H abundance using nuclear emission-line ratios produce reliable results, which are in agreement with the O/H values obtained from the independent method of galactic metallicity gradient extrapolation.  Finally, the results from the WHAN diagram combined with the fact that the high  excitation level of the gas has to be maintained at kpc scales,   we suggest that the main ionizing source of the UGC\,4805 nucleus probably has a stellar origin rather than an AGN.

\end{abstract}



\section{Introduction}

Determinations of the chemical abundances of Active Galactic Nuclei (AGNs) and Star-Forming regions (SFs) are essential for understanding the chemical evolution of galaxies and, consequently,  of the Universe.

Among the heavy elements present in the gas phase of AGNs and SFs (e.g., O, N, S), oxygen is the element with more accurate abundance determinations. This  is because prominent emission-lines from the main ionic stages of oxygen can be easily detected in the optical spectra of these objects, making it a good tracer
of the metallicity (e.g., \citealt{2003ApJ...591..801K, 2008MNRAS.383..209H, 2015MNRAS.453.4102D, 2020MNRAS.492..468D}).
Therefore, hereafter we use  metallicity ($Z$) and oxygen abundance [12 + $\log$(O/H)] interchangeably.
 Abundance estimations based on the direct method, also known as $T_{\rm e}$-method, are commonly used to determine chemical abundances of
 gas phase of SFs  (for a review see \citealt{2017PASP..129h2001P, 2017PASP..129d3001P}). These estimations  seem to be more reliable than those derived using empirical or theoretical relations between the different electron temperatures \citep{2006MNRAS.372..293H, 2008MNRAS.383..209H}.
  In fact, the compatibility  between
oxygen abundances in nebulae located in the  solar neighborhood and the ones derived from observations
of the weak interstellar \ion{O}{i}$\lambda$1356 line towards the stars (see \citealt{2003A&A...399.1003P} 
and references therein) sustains the accuracy of the $T_{\rm e}$-method.
 This method is based on determinations of nebular electron temperatures, which requires measurements
of auroral emission-lines, such as [\ion{O}{III}]$\lambda$ 4363 and [\ion{N}{II}]$\lambda$ 5755,  generally weak (about 100 times weaker than H$\beta$) or not measurable
in objects with high metallicity and/or low excitation (e.g., \citealt{1998AJ....116.2805V, 2007MNRAS.382..251D}). In the cases
that auroral lines can not be measured, indirect or strong-line methods can be used to estimate the oxygen
abundance, as proposed by \citet{1976ApJ...209..748J} and \cite{10.1093/mnras/189.1.95}. This method is based on calibrations between the oxygen abundance or metallicity and strong emission-lines, easily measured in SF spectra (for a review see \citealt{2010A&A...517A..85L, 2019A&ARv..27....3M, 2019ARA&A..57..511K}).

For AGNs,  chemical abundance determinations are preferably carried out
in Narrow Line Regions (NLRs) of Seyfert 2 nuclei due to the relatively low velocity  ($v \: \la \: 400 \: \rm km  \: s^{-1}$, \citealt{2017MNRAS.469.3125C}) of the shock waves present in the gas and
their low electron density ($N_{\rm e} \: \la \: 2000 \: \rm cm^{-3}$, \citealt{2013MNRAS.430.2605Z, 2014MNRAS.443.1291D}; for a review see \citealt{2020MNRAS.492..468D}).
Oxygen abundance estimations for  NLRs of Seyfert 2 have been obtained by using the
$T_{\rm e}$-method (\citealt{1992A&A...266..117A, 2008ApJ...687..133I, 2015MNRAS.453.4102D, 2020MNRAS.492..468D})
and strong-line methods (e.g., \citealt{Storchi_Bergmann_1998, 10.1093/mnras/stx150, 2020MNRAS.492.5675C}).
Studies based on strong-line methods
  have indicated that Seyfert 2 nuclei in the local universe ($z \: < \: 0.4$) present similar metallicity (or abundances) as those in metal rich \ion{H}{ii} regions, i.e., no extraordinary enrichment has been observed in AGNs, with these objects exhibiting  solar
or slightly over-solar metallicities. This result agrees with predictions of chemical evolution
models for spiral and elliptical galaxies (e.g., \citealt{molla}).

An opposite situation is found for Low-Ionization Nuclear Emission-line Regions (LINERs), whose chemical abundance studies are rare in the literature. This class of objects appear in 1/3 of galaxies in the local universe \citep{netzer_2013}, and their ionization sources are still an open problem in astronomy.  \cite{1980A&A....87..152H} suggested that these nuclei have  gas shocks as their main ionization/heating source. Later,  \cite{1983ApJ...264..105F} proposed that LINERs could be ionized by accretion gas into a central black hole (AGN) but with lower ionization parameters \textit{(U)} than those found in Seyferts. Therefore, the difference between LINERs and other AGN types would consist of the order of the ionization parameter \citep{1993ApJ...417...63H}. However, \cite{10.1093/mnras/213.4.841} and \cite{1992ApJ...399L..27S} proposed a new ionization model, i.e., LINERs are ionized by hot stars, but contrary to SFs, they are old stars (0.1-0.5 Gyr) that came out from the main sequence (e.g., in the post-Asymptotic Giant Branch, p-AGB).
Based on this scenario, \cite{Taniguchi_2000} showed that
photoionization models considering Planetary Nebula Nuclei (PNNs) with a temperature of $10^{5}$ K as ionizing sources can reproduce the region occupied, at least, by a subset of type 2 LINERs in optical emission-line ratio diagnostic diagrams. \cite{2014arXiv1409.2966W} found that these objects have composite ionizing sources, i.e., more than one mechanism is responsible for the ionization of the gas. This explanation was also proposed by \cite{Yan_2012}, \cite{ refId0}, and \cite{2013A&A...558A..34B}. 

The unknown nature of the ionizing sources and excitation mechanisms of LINERs hinder determination of their  metallicity using the  $T_{\rm e}$-method and/or strong-line methods (e.g., \citealt{Storchi_Bergmann_1998}). \citet{2010A&A...519A..40A} analysed intermediate-resolution optical spectra of a sample of LINERs and derived oxygen abundances considering these objects as AGNs (by using the
\citealt{Storchi_Bergmann_1998} calibrations) and as SFs (by using the
\citealt{1999ApJ...514..544K} calibration). These authors found that when AGNs are assumed as ionizing sources, higher oxygen values are derived than for those assuming hot stars, which provide sub-solar abundances. On the other hand, oxygen abundance estimations based on the extrapolation  of  disk abundance gradients to the central part of the galaxies  (an independent method)   by \citet{2012A&A...543A.150F} indicate over-solar oxygen abundances for three LINERs (NGC\,2681, NGC\,4314, and NGC\,4394). 

Recently,  semi-empirical calibrations between the oxygen abundance (or metallicity) and strong-emission lines of Seyfert 2 were obtained by \citet{10.1093/mnras/stx150} and \citet{2020MNRAS.492.5675C}. In addition,  several
methods to determine the oxygen abundance gradients in spiral galaxies are available in the literature
\citep[see][]{1992MNRAS.259..121V,1994ApJ...420...87Z,1998AJ....116.2805V,2004A&A...425..849P,Pilyugin+07,lopez10}.  These methods, together with data from the Mapping Nearby Galaxies at the Apache Point Observatory \citep[MaNGA, ][]{2015ApJ...798....7B}, offer a powerful opportunity to determine  the chemical abundances of LINERs and to produce insights about the ionization mechanisms of these objects.

In previous papers, we have analysed oxygen abundance in Seyfert 2 nuclei using the $T_{\rm e}$-method, photoionization model grids, and {\sc HCM} code (see \citealt{2014MNRAS.443.1291D,10.1093/mnras/stx150,2019MNRAS.489.2652P,2019MNRAS.486.5853D,2020MNRAS.492.5675C,2020MNRAS.492..468D,2020MNRAS.496.3209D}). Although, the semi-empirical calibrations between metallicity and strong-emission lines of Seyfert 2 obtained by \citealt{10.1093/mnras/stx150} and \citealt{2020MNRAS.492.5675C} along with the AGN photoionization model grids and SF calibrations, are applied in this paper,  the object class studied here  and the methodology applied are also different. The main goal of this work is to determine the oxygen abundance in relation to the hydrogen abundance (O/H) in the central region of the LINER galaxy 
UGC\,4805 (redshift $z=0.02698$), in combination with data from the Mapping Nearby Galaxies at the Apache Point Observatory (MaNGA, \cite{2015ApJ...798....7B}.
We assumed a spatially flat cosmology with
 $H_{0}$\,=\,71 $ \rm km\:s^{-1} Mpc^{-1}$, $\Omega_{m}=0.270$, and $\Omega_{\rm vac}=0.730$  \citep{2006PASP..118.1711W}, which leads to a spatial scale of 0.535 kpc/arcsec at the distance of UGC\,4805.
This paper is organized as follows: in Section~\ref{dataobs} the observational data of UGC\,4805
are described; Section~\ref{meth} contains 
the methodology used to estimate the oxygen abundance of the nucleus and along the disk of UGC\,4805; 
in Section~\ref{res}, the results for the nuclear oxygen abundance are given;  while discussion and conclusions
of the outcome are presented in Sections~\ref{disc}  and ~\ref{conc}, respectively.

\section{Data}
\label{dataobs}

\begin{figure*}
\centering
\begin{minipage}{.49\textwidth}
\includegraphics[width=0.7\textwidth]{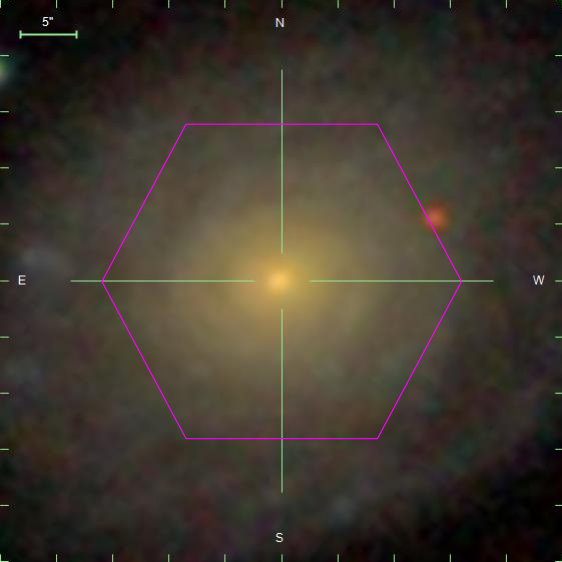}
\end{minipage}
\begin{minipage}{.49\textwidth}
\includegraphics[width=1.1\textwidth]{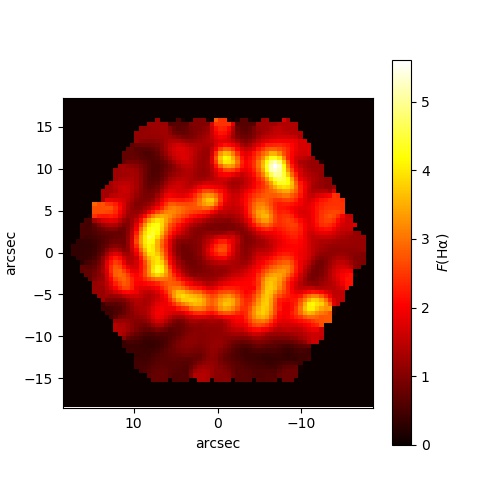}
\end{minipage}\caption{Left panel: SDSS false colour image combining the $gri$ bands of UGC\,4805 taken from the  MaNGA survey \citep{2017AJ....154...28B}. The IFU field of view is indicated in purple. Right panel: Map of the H$\alpha$ flux (in units of $10^{-17}$ erg/cm$^{2}$/spaxel). }
\label{figure1}
\end{figure*}

\begin{figure*}
\centering
\includegraphics[width=1\textwidth]{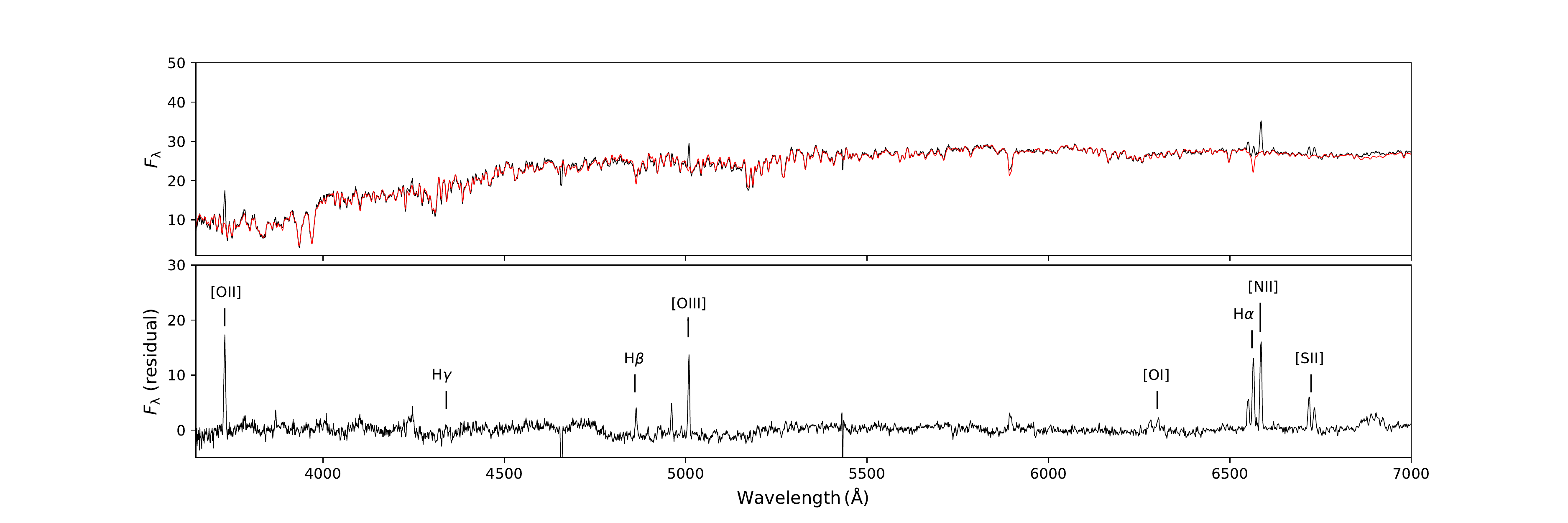}
\caption{Upper panel: Stellar population synthesis for the nuclear region of UGC\,4805
within a  circular aperture with a  radius equal to 1.5 arcsec ($\sim$1 Kpc).
Observed and synthetic spectra are in black and red, respectively. 
 Lower Panel: Pure emission spectrum of the UGC\,4805 nucleus.
Emission lines are identified in the plot. The flux is in units of $10^{-15} \rm{ergs}\, \rm{cm^{-2}}\, \rm{s^{-1}}\, \AA$.}
\label{field_ha}
\end{figure*}

MaNGA survey is an Integral Field Spectroscopy (IFS) survey\footnote{sdss.org/surveys/manga/} 
\citep{2017AJ....154...28B} developed to observe about 10\,000 galaxies until 2020 using Integral Field Units (IFUs).  This survey is part of the fourth version of the Sloan Digital Sky Survey (SDSS-IV, \citealt{2017AJ....154...28B}) and utilises the 2.5 m Sloan Telescope in its spectroscopic mode.
The spectra have a wavelength coverage of 3\,600 - 10\,300 \AA, with a spectral resolution of $R \sim$ 1\,400  
at $\lambda\sim$4\,000  \AA\  and $R \sim$ 2\,600 at $\lambda\sim$9\,000 \AA. 
The angular size of each spaxel is 0.5 arcsec, and the average Full Width Half Maximum  (FWHM) of the
MaNGA data is 2.5 arcsec. For details about the strategy of observations and data reduction see  \citet{2015AJ....150...19L} and \citet{2016AJ....152...83L}, respectively.
From the objects whose data are available in the MaNGA survey, we selected those presenting LINER nuclei and  disk emission, preferably from objects classified as SFs. Based on these selection criteria, we selected 81 objects. In this work, we present a detailed analysis of the spiral galaxy UGC\,4805, an object with a classical LINER nuclear emission and with the largest number of star-forming emission spaxels along the disk.
The spectrum of each spaxel was processed according to the steps listed below:

\begin{itemize}
 \item  To obtain the nebular spectrum of each spaxel not contaminated by the stellar population continuum, i.e., the pure nebular spectrum,  we use the stellar population synthesis code  {\scriptsize\,STARLIGHT} developed by  
\citet{2005MNRAS.358..363C, 2006MNRAS.370..721M, 2007MNRAS.381..263A}. This code fits the observed spectrum of a galaxy using a combination of Simple Stellar Populations\,(SSPs), in different proportions and excluding the emission lines. 
We use a spectral basis of  45 synthetic SSP spectra with three metallicities $Z$\,=\,0.004, 0.02  ($Z_{\odot}$), and 0.05, 
and 15 ages from 1 Myr up to 13 Gyr, taken from the evolutionary synthesis models of \citet{2003MNRAS.344.1000B}. The reddening law by \citet{1989ApJ...345..245C} was considered.
The stellar spectra of the SSPs were convolved with an elliptical Gaussian function to achieve
the same spectral resolution as the observational data and transformed to the rest frame.\\

\item The emission lines are fitted with Gaussian profiles. For more details about the synthesis method and the fitting of emission lines, see \citet{2016MNRAS.462.2715Z}.  \\

\item The residual extinction associated with the gaseous component for each
spatial bin was calculated by comparing the observed value of the
H$\alpha$/H$\beta$ ratio to the theoretical value of 2.86 obtained by \cite{1987MNRAS.224..801H} for an electron temperature of 10\,000 K and an electron density of 100 cm$^{-3}$. 
\end{itemize}
Fig.~\ref{figure1} presents the  SDSS false colour image obtained combining the $gri$ bands of UGC\,4805 and the resulting 2D map of the H$\alpha$ flux. Observe the very separated and clear nucleus and a bright star-forming ring in the disk
at $\sim$8 arcsec ($\sim 4.2$ kpc). In Fig.~\ref{field_ha} (top panel), the observed (in black) and synthetic (in red) spectra
of the central region of UGC\,4805 are shown. Fig.~\ref{field_ha} (bottom panel) also presents the pure emission spectrum, i.e., after the SSP subtraction, as well as emission line identifications. The nuclear emission was obtained by integrating the flux of the central region considering a radius of 1.5 arcsec ($\sim$1 kpc), which corresponds approximately to the mean value of the seeing during the observations. In Table~\ref{fluxos} the reddening corrected emission-line intensities (in relation to H$\beta$=100), 
the reddening function $f(\lambda)$,  the logarithmic extinction  coefficient $c$(H$\beta$), the visual extinction A$_{\rm V}$,  and the equivalent width  of H$\alpha$ [$\rm W_{H\alpha}$] 
of the LINER nucleus of UGC\,4805 are listed. 
The H$\beta$ luminosity (in units of erg/s) was  also calculated and listed in Table~\ref{fluxos}, considering a distance of 119 Mpc.

\begin{figure*}
\includegraphics*[angle=0,width=0.9\textwidth]{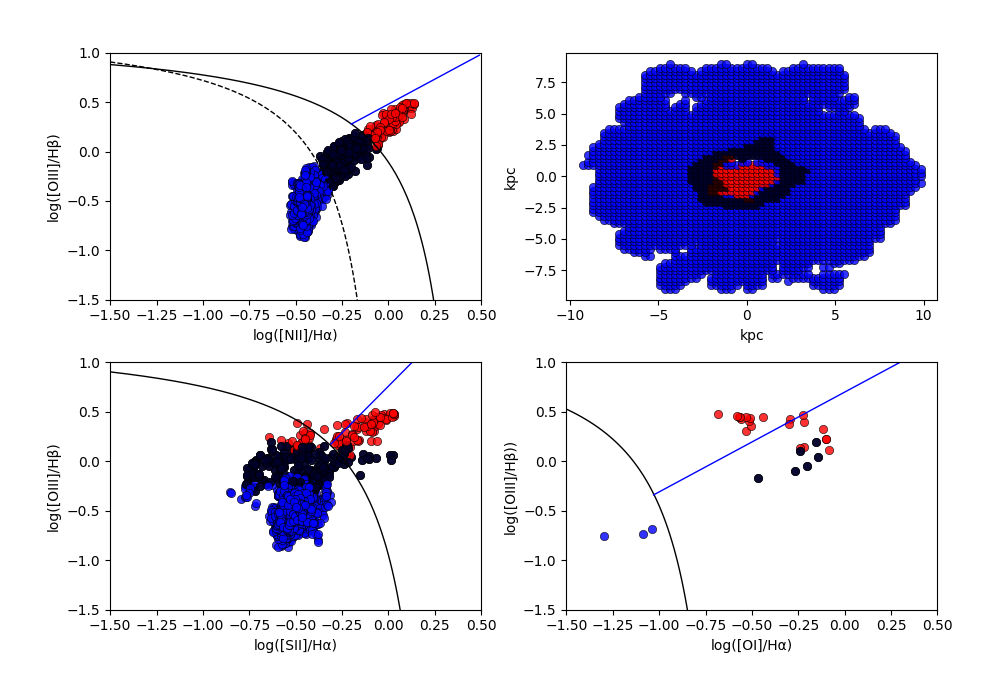}
\caption{Top left panel:  $\log([\ion{O}{iii}] \lambda 5007/\rm H\beta)$ versus $\log([\ion{N}{ii}]$ $\lambda 6584/\rm H\alpha)$  
diagnostic diagram. Black solid curve represents the theoretical upper limit for the star-forming regions proposed by \citealt{kewley01} (Ke01),  the black dashed curve is the  empirical star-forming limit proposed by \citet{2003MNRAS.346.1055K} (Ka03), and the blue solid line represents the separation  between  Seyferts and LINERs from  \citet{2006MNRAS.372..961K} (Ke06). The region between the Ke01 and Ka03 lines is denominated the composite region (black points). Top right panel: Distribution of the UGC\,4805 regions 
accordingly to their main excitation mechanism as showed 
in the  $\log([\ion{O}{iii}] \lambda 5007/\rm H\beta)$ versus $\log([\ion{N}{ii}]$ $\lambda 6584/\rm H\alpha)$  diagram (top left panel). Bottom left panel: $\log( [\ion{O}{iii}] \lambda 5007/\rm H\beta)$ versus  $\log([\ion{S}{ii}](\lambda\lambda6716+31)/\rm H\alpha$) diagram.  Bottom right panel: $\log([\ion{O}{iii}]\lambda 5007/\rm H\beta)$ versus $\log([\ion{O}{i}] \lambda 6300/\rm H\alpha)$ diagram. Red points represent the AGN-like spaxels and  blue points  the SF-like spaxels of UGC\,4805, according to $\log([\ion{O}{iii}] \lambda 5007/\rm H\beta)$ versus $\log([\ion{N}{ii}]$ $\lambda 6584/\rm H\alpha)$ diagram. }
\label{bpt_diag}
\end{figure*}

\begin{figure} 
\includegraphics[angle=0,width=0.45\textwidth]{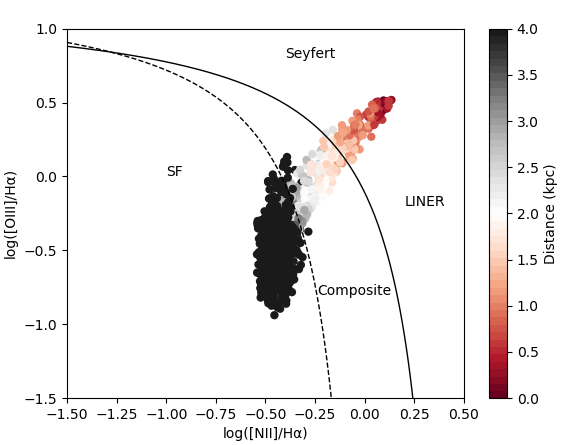}
\caption{$\log([\ion{O}{iii}] \lambda 5007/\rm H\beta)$ versus $\log([\ion{N}{ii}]$ $\lambda 6584/\rm H\alpha)$  diagnostic diagram. The colour of each point corresponds to its distance from the galaxy centre, with the reddest points representing the central spaxels.}
\label{bpt_center}
\end{figure}

 The identification of the dominant ionization mechanism of the emitting gas across the galaxy is essential to determine chemical abundances. To do that, we used the $[\ion{O}{iii}]\lambda 5007/\rm H\beta$ versus $[\ion{N}{ii}]\lambda 6584/\rm H\alpha$,   $[\ion{O}{iii}] \lambda 5007/\rm H \beta$ versus  $[\ion{S}{ii}](\lambda \lambda 6716+31)/\rm H \alpha$, and $[\ion{O}{iii}] \lambda 5007/\rm H\beta$ versus $[\ion{O}{i}]\lambda 6300/\rm H \alpha$ diagnostic diagrams proposed by \citet{1981PASP...93....5B},  commonly known as BPT  diagrams, to classify each spaxel of UGC\,4805.  The empirical and theoretical criteria proposed by \citet{kewley01} and \citet{2003MNRAS.346.1055K}, respectively, were considered to classify objects in \ion{H}{ii}-like regions,  composite, and AGN-like objects. Furthermore, the separation between  Seyferts and LINERs  proposed by \cite{2006MNRAS.372..961K} was used.  Fig.~\ref{bpt_diag} shows these BPT diagrams for each spaxel of UGC\,4805 and the distribution of the regions in the galaxy 
according to  $[\ion{O}{iii}]/\rm H\beta$ versus 
$[\ion{N}{ii}]/\rm H\alpha$ diagram. As can be seen in these diagrams, the central area of the galaxy is classified as LINER. Fig. ~\ref{bpt_center} shows the same $[\ion{O}{iii}]\lambda 5007/\rm H\beta$ versus $[\ion{N}{ii}]\lambda 6584/\rm H\alpha$ diagram as Fig.~\ref{bpt_diag} (top left panel), but as a function of the distance to the centre of the galaxy. The colour  of each point corresponds to its distance from the galaxy centre, with the reddest points representing the central spaxels. As can be noted in this figure, the points closest to the centre lie in the LINER region. In Addition,  the distance to the galaxy centre and the location in the diagram are connected, so that the points that approach  the centre of the galaxy moves away  from the line that separates SF-like objects from AGN-like ones.

On the other hand, the diagram introduced by \citet{2011MNRAS.413.1687C} uses the equivalent width of H$\alpha$ ($\rm W_{H\alpha}$) and is known as a WHAN diagram.   This diagram can to discriminate genuine low-ionization AGNs from galaxies that are ionized by  evolved low-mass stars, i.e. the post-Asymptotic Giant Branch (post-AGB).  The WHAN diagram identifies 5 classes of galaxies, namely:
\begin{enumerate}
  \item Pure star forming galaxies: $\log(\ion{N}{ii}/\rm H\alpha) \: < \: -0.4$ and  $\rm W_{H\alpha} \: > \: 3$ \AA.
  \item Strong AGN (i.e., Seyferts): $\log(\ion{N}{ii}/\rm H\alpha)\: > \: -0.4$ and  $\rm W_{H\alpha} \: > \:  6$ \AA.
  \item Weak AGN: $\log(\ion{N}{ii}/\rm H\alpha) \:  > \:  -0.4$ and  $\rm W_{H\alpha}$ between 3 and 6 \AA.
  \item Retired galaxies (i.e., fake AGN): $\rm W_{H\alpha} \: < \: 3$ \AA.
  \item Passive galaxies (actually, line-less galaxies): $\rm W_{H\alpha}$ and $\rm W_{\ion{N}{ii}} \: < \: 0.5$ \AA.
\end{enumerate}
According to this classification, the UGC\,4805 nucleus is a Retired Galaxy and, thus, it is  ionized by post-AGB stars.


\begin{table}
\centering
\caption{Reddening corrected emission-line intensities (in relation to H$\beta$=100),  reddening function $f(\lambda)$,  the logarithmic extinction  coefficient $c$(H$\beta$),  the visual extinction A$_{\rm V}$, and the  H$\beta$ luminosity (erg/s)
of the UGC\,4805 nucleus. The estimations were obtained considering a radius of 1 kpc.}
\label{fluxos}
\begin{tabular}{@{}lrr@{}}
\hline

                & $f(\lambda$) &  Measurements \\
\hline
[\ion{O}{II}]\,$\lambda$3727  & 0.33 & 327 $\pm$ 5 \\

[\ion{O}{III}]\,$\lambda$4959 & $-$0.02 & 91 $\pm$ 2 \\

H$\beta$\,$\lambda$4861       & 0.00 & 100 $\pm$ 3 \\

[\ion{O}{III}]\,$\lambda$5007 & $-$0.04  & 242 $\pm$ 3\\

[\ion{N}{II}]\,$\lambda$6548  & $-$0.35 &  126 $\pm$ 3\\

 [\ion{O}{I}]\,$\lambda$6300   &  $-$0.29 &  21 $\pm$ 5 \\

H$\alpha$\,$\lambda$6563      & $-$0.35 & 286 $\pm$ 3\\

[\ion{N}{II}]\,$\lambda$6584  &$-$0.35  & 321 $\pm$ 4 \\

[\ion{S}{II}]\,$\lambda$6717  & $-$0.36  & 135 $\pm$ 3 \\

[\ion{S}{II}]\,$\lambda$6731   & $-$0.37 & 96 $\pm$ 3 \\

$c$(H$\beta$) & --- & 0.19  $\pm$ 0.005\\

W$_{\rm H\alpha}$ &  --- &  1.65 $\pm$ 0.21 [\AA] \\

 A$_{\rm V}$ &  --- &  0.37 [mag] \\

 $\log$[$L$(H$\beta$)]  &  --- &  38.86 [erg/s]\\[2pt]

\hline

\end{tabular}
\end{table}

\section{Oxygen abundance determination}
\label{meth}

To obtain the oxygen abundance of the UGC\,4805 nucleus, five calibrations
of SFs were used  to extrapolate the radial oxygen abundance for  the central region. This method has been used
by several authors (e.g., \citealt{1992MNRAS.259..121V,1998AJ....116.2805V,2004A&A...425..849P, 2019MNRAS.483.1901Z}) and it produces an independent estimation of the oxygen abundance of nuclear regions. 
Recently, \citet{2020arXiv200205744M} measured gas-phase metallicity, ionisation parameter and dust extinction for a representative sample of 1795 local star-forming galaxies using integral field spectroscopy from the SDSS-IV MaNGA survey, showing the extensive reliability of this survey in this type of study. 
In addition, calibrations between the gas phase O/H abundance (or metallicity) and strong emission-lines for Seyfert 2 AGNs and  photoionization model results were considered to estimate the  UGC\,4805 nucleus oxygen abundance. Each 
 method is described below.

\subsection{Star-forming regions}
\label{hiical}

The goal of this work is to determine the oxygen abundance in the nuclear region of UGC\,4805.  In principle,  determinations of oxygen abundances based on measurements of temperature sensitive line ratios, for example $[\ion{O}{iii}]\lambda\,4363$ and $[\ion{N}{ii}]\lambda\,5755$, should provide  more accurate estimates of O/H \citep{2003ApJ...591..801K}, because these are free from the  uncertainties of photoionization models (e.g., \citealt{2002RMxAC..12..219V, 2003ApJ...591..801K}), considered in the majority of strong-line methods (e.g., \citealt{2002ApJS..142...35K}). 
Unfortunately, electron temperature sensitive line ratios were not measured in the UGC\,4805  spectra. In these cases, only strong-line methods would be used  to determine the oxygen abundances in the $\ion{H}{ii}$ regions along the UGC\,4805 disk and, then, to obtain the central intersect  O/H abundance. The strong-line methods considered in this  work 
to derive the O/H gradient are briefly described below.

\begin{itemize}
\item \cite{edmunds84}: This theoretical calibration, obtained by using the model calculations
by \citet{1980ApJ...236..119D} and \citet{1980MNRAS.193..219P}, is based on the  $R_{23}$=([\ion{O}{ii}]$\lambda$3727+[\ion{O}{iii}]$\lambda\lambda$4959+5007)/H$\beta$ index and the equations are given by
\begin{equation}{\label{edeq1}}
12 +\log {\rm(O/H)_{up}} = 8.76 - 0.69 \log R_{23}
\end{equation}
and
\begin{equation}{\label{edeq2}}
  12 + \log {\rm (O/H)_{low}} = 6.43 + 1.67 \log R_{23},
\end{equation}
where "up" and "low" mean the equations for the upper and lower branch of the (O/H)-$R_{23}$  calibration, respectively.  \\

\item \cite{denicolo}: These authors proposed a calibration between the O/H abundance
and the $N2=\log$([\ion{N}{ii}]$\lambda$6584/H$\alpha$) line ratio, originally proposed by \citet{1994ApJ...429..572S} as a metallicity indicator for \ion{H}{ii} regions. For the low metallicity regime ($\rm 12 + \log(O/H) \: < \: 8.4$), \cite{denicolo} considered O/H values calculated through the $T_{\rm e}$-method and for the high metallicity regime
abundance estimations based on calibrations by \citet{1991ApJ...380..140M} and \citet{2000MNRAS.312..130D}. The expression proposed by \cite{denicolo} is 
\begin{align*}
 12 + \mathrm{\log(O/H)} & = 9.12 + 0.73 \times N2. 
\end{align*}
This calibration is valid for the range  of $7.2 < 12 + \mathrm{\log(O/H)} < 9.1$.\\

\item \cite{2004MNRAS.348L..59P}: These authors used a sample of extragalactic $\ion{H}{ii}$ regions and the 
$O3N2 = \log \left(\frac{[\mathrm{OIII}]\lambda\,5007/ \mathrm{H}\beta}{[\mathrm{NII}]\lambda\,6583/\mathrm{H}\alpha}\right)$ parameter to derive the calibration:
\begin{align*}
 12 + \mathrm{\log(O/H)} & = 8.73 - 0.32 \times \textit{O3N2},
\end{align*}
valid for the range  of $8.0 < 12 + \mathrm{\log(O/H)} < 9.0$. \cite{2004MNRAS.348L..59P} considered O/H values calculated using the $T_{\rm e}$-method for most cases and a few estimations based on detailed photoionization models.\\

\item \cite{dors}: These authors built photoionization model sequences
to reproduce the emission-line ratio intensities of \ion{H}{ii} regions located
along the disks of a sample of spiral galaxies to derive O/H gradients. \cite{dors} obtained the semi-empirical calibration 
\begin{align*}
 12 + \mathrm{\log(O/H)} & = 8.96 - 0.03 x - 0.1 x^{2} -  0.21 x^{3} - 0.26 x^{4},
\end{align*}
with $x = \log{R_{23}}$. This calibration is
valid for the upper branch of the (O/H)-$R_{23}$ relation (i.e.,  $12 + \log({\rm O/H}) \: > \: 8.2$). \\

\item \cite{pilyugin16}: These authors used a sample of \ion{H}{II} regions with abundances determined by the `counterpart' method ($C$ method) 
to derive a calibration based on oxygen and nitrogen emission lines. These empirical calibrations use the excitation parameter 
$P = R_{3}/(R_{2} + R_{3})$, and  $N2$,  
where $R_2$ = [\ion{O}{II}]($\lambda\,$3726 +  $\lambda\,$3729)/H$\beta$  and
$R_{3}$ = [\ion{O}{III}]($\lambda$\,4959 + $\lambda$\,500 7)/H$\beta$.

Two equations were obtained,  one for \ion{H}{II} regions with $N2 \lid -0.6$  (the lower branch), defined by
\begin{align*}
12 + \log (\mathrm{O/H}) &= 7.932 + 0.944 \times \log (R_3/R_2)+0,695  \times  N2 + \\
                         & \quad + (0.970 - 0.291 \times \log (R_3/R_2) + \\ & \quad - 0.019 \times  N2 ) \times \log R_2,
\end{align*}
and another for $N2 \: \gid \: -0.6$ (the upper branch), where  the following equation is valid
\begin{align*}
12 + \log (\mathrm{O/H}) &= 8.589 + 0.022 \times \log (R_3/R_2)+0.399 \times  N2 + \\  
 & \quad  + (-0.137 + 0.164 \times  \log (R_3/R_2) + \\  
 & \quad  +  0.589\times  N2 )   \times \log R_2.
\end{align*}
 
\end{itemize}
This method is similar to the $C_{NS}$ method proposed by  \citet{2012MNRAS.424.2316P} and it yields
O/H abundance  values similar to those derived through the $T_{\rm e}$-method.

\subsection{Active Galactic Nuclei}
\label{agnc}

\begin{itemize} 
\item \cite{Storchi_Bergmann_1998}: The first calibrations between the oxygen abundance and strong narrow emission-line ratios of AGNs were the theoretical ones proposed by \cite{Storchi_Bergmann_1998}. These authors used photoionization model sequences,
built with the {\sc Cloudy} code \citep{1996hbic.book.....F},  and proposed the calibrations

\begin{align*}
\begin{matrix}
  (\mathrm{O/H}) &=  8.34 + 0.212x - 0.012 x^{2} - 0.002 y+0.007xy +  \\
                 & \quad - 0.002x^{2}y +  6.52 \times 10^{-4}y^{2}+2.27 \times 10^{-4}xy^{2}+ \\
  & \quad + 8.87 \times 10^{-5}x^{2}y^{2},
   \end{matrix}
\label{sb1}
\end{align*}
\begin{align*}
\begin{matrix}
  (\mathrm{O/H}) &= 8.643 -0.275u + 0.164 u^{2} + 0.655 v - 0.154 uv  + \\
  & \quad - 0.021u^{2}v + 0.288v^{2} + 0.162uv^{2} + 0.0353u^{2}v^{2},
\end{matrix}
\end{align*}
where $x$ = [\ion{N}{II}]($\lambda \lambda$6548,6584)/H$\alpha$, $y =$ [\ion{O}{III}]($\lambda \lambda$4949,5007)/H$\beta$,
 $ u = \log $([\ion{O}{II}]($\lambda \lambda$3726,3729)/[\ion{O}{III}]($\lambda \lambda$4959,5007),   and  $v = \log$ ([\ion{N}{II}]($\lambda \lambda$6548,6584)/H$\alpha)$.

These calibrations are valid for the range  of  $8.4 \: < 12 + \: \mathrm{\log(O/H)} \: \:< 9.1$. Differences
between O/H estimations derived using these calibrations are in of order of 0.1 dex \citep{Storchi_Bergmann_1998, 2010A&A...519A..40A, 2020MNRAS.492..468D,2015MNRAS.453.4102D}.  For  LINERs, \cite{Storchi_Bergmann_1998}  found that the calibrations above yield lower values than those derived from the  extrapolation of  O/H abundance gradients, suggesting that the assumptions of their models are not representative for LINERs. It should be mentioned that they indicated  that their sample of LINERs was too small (only four objects) to provide a firm conclusion about the application of their method to this kind of object. They also suggest that a more extensive sample needs to be used to test their  calibrations.\\

\item \cite{10.1093/mnras/stx150} proposed a semi-empirical calibration between the metallicity and the \textit{N2O2}$ = \log ([\mathrm{\ion{N}{II}}] \lambda\,6584/[\mathrm{\ion{O}{II}}]\lambda\,3727)$ index. The calibration derived by these authors  
was obtained upon a comparison between observational and photoionization model predictions
of the [\ion{O}{iii}]$\lambda$5007/[\ion{O}{ii}]$\lambda$3727 versus $N2O2$ line ratios and given by
\begin{align*}
   (Z/{\rm Z_{\odot}}) &= 1.08 (\pm 0.19) \times N2O2^{2} + 1.78 (\pm 0.07) \times N2O2 + \\
   & \quad 1.24 (\pm 0.01).
\end{align*}
This calibration is valid for 
 $-1.4 \: \la \: ([\ion{O}{iii}]/[\ion{O}{ii}])\: \la \: 2$
 and $-1.0 \: \la \: N2O2\: \la \: 1$.\\

\item \citet{2020MNRAS.492.5675C} used the same methodology as
\cite{10.1093/mnras/stx150} to calibrate NLRs metallicities of Seyfert 2
nuclei with the $N2$  emission-line ratio. This
ratio is practically independent of the flux calibration and reddening correction.
These authors proposed the following calibration
\begin{equation}
\label{eq1}
(Z/Z_{\odot})=a^{N2}+b,
\end{equation}
where $a=4.01\pm0.08$ and $b=-0.07\pm0.01$. This calibration is valid
for $-1.4 \: \la \: ([\ion{O}{iii}]/[\ion{O}{ii}])\: \la \: 2$
and $-0.7 \: \la \: (N2) \: \la \: 0.6$. \citet{2020MNRAS.492.5675C}
also proposed a relation between  the ionization parameter ($U$) and the  [\ion{O}{iii}]$\lambda$5007/[\ion{O}{ii}]$\lambda$3727 line ratio, almost independent of other nebular parameters, and given by
\begin{equation}
\label{eq2}
\log U=(0.57\pm 0.01 \: x^{2})+(1.38\pm 0.01 \: x) - (3.14\pm 0.01),
\end{equation}
where $x=\log$([\ion{O}{iii}]$\lambda$5007/[\ion{O}{ii}]$\lambda$3727).
\end{itemize}  

Although the  AGN calibrations  above were developed 
for Seyfert 2 nuclei, in this paper, we consider them to derive the O/H abundance in the LINER nucleus of UGC\,4805, and we compared the resulting   values to those derived from extrapolation  of  oxygen abundance gradients for central parts of this galaxy.

\subsection{Photoionization models}
\label{modd}

\begin{figure}
\centering
\includegraphics*[angle=-90,width=\columnwidth]{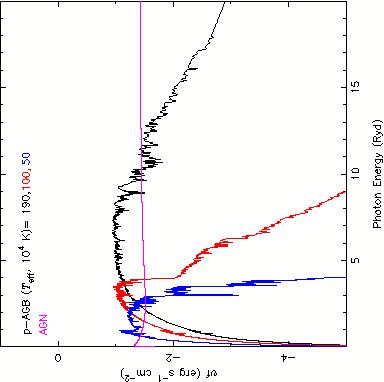}
\caption{Comparison between the p-AGB star and AGN SEDs assumed the ionizing source in the photoionization models. The atmosphere models by \citet{2003A&A...403..709R} and three effective temperature values (as indicated) are considered. The AGN SED is represented by a multi-component continuum with spectral index $\alpha_{ox}=-1.0$ (see Eq.~\ref{apxeq}).}
\label{fsed}
\end{figure}

To reproduce the observed line ratios  of UGC\,4805 LINER nucleus with the goal of deriving the O/H abundance and
the ionization parameter ($U$), we built photoionization model grids using version 17.00 of the {\sc CLOUDY} code \citep{2017RMxAA..53..385F}. These models are similar to the ones used in \citet{2020MNRAS.492.5675C}, 
and a brief description of the input parameters is presented below:

\begin{enumerate}
    \item SED:  The models consider two  distinct Spectral Energy Distributions (SEDs): one
       to represent an AGN and another representing p-AGB stars. 
       The AGN SED is a multi-component continuum, similar to that observed in typical AGNs. As described in the Hazy manual of the Cloudy code \footnote{\url{http://web.physics.ucsb.edu/~phys233/w2014/hazy1_c13.pdf}}, it is composed by the sum of two components. 
       The first one is a Big Bump component peaking at $\approx$ 1 Ryd, parametrized by the temperature of the bump, assumed to be $5 \: \times \: 10^{5}$ K, with a high-energy exponential cutoff and an infrared
exponential cutoff at 0.01 Ryd. The second component is  an X-ray power law with spectral index $\alpha_x=-1$ that is only added for
energies greater than 0.1 Ryd to prevent it from extending into the infrared.  The X-ray power law is not extrapolated below 1.36 eV or above 100 keV: for energies lower than 1.36 eV it is set to zero (since the bump dominates for these energies), and for energies above 100 keV,  the continuum falls off as $\nu^{-2}$.
The  $\alpha_{ox}$ spectral index defined as the slope of a power law between 2\: keV and 2500\:\AA\ is the parameter that provides the normalization of the X-ray power law to make it compatible with the thermal component. It is given by 
      \begin{equation}
      \label{apxeq}
\alpha_{ox}= \frac{\log [F(2\: {\rm keV})/F(2500\: \textrm{\AA})]}{\log [\nu(2 \: {\rm keV})/\nu(2500 \: \textrm{\AA})]}, 
\end{equation}
where $F$ is the flux at 2\:keV,  2500\:\AA\ and $\nu$ are the corresponding frequencies \citep{1979ApJ...234L...9T}.
This AGN SED  generates a continuum similar to that used by  \citet{1997ApJS..108..401K}.
In all our AGN models, a fixed value of $\alpha_{ox}=-1.0$ is assumed.  \citet{2020MNRAS.492.5675C}
found that models with $\alpha_{ox} \: \la \: -1.0$ trend not to reproduce optical emission
line ratios of Seyfert 2 nuclei (see also \citealt{2017MNRAS.468L.113D, 2019MNRAS.489.2652P}).
Moreover, $\alpha_{ox}\sim-1.0$ has been derived in observational studies of
LINERs and low luminosity AGNs (see \citealt{ho99, 2010ApJS..187..135E, 2007MNRAS.377.1696M, 2012A&A...539A.104Y}).

In the case of the stellar SED, we consider p-AGB stars atmosphere models by \citet{2003A&A...403..709R}
assuming the available values for the effective temperatures: 
 $T_{\rm eff}= 50, 100$,  and 190 kK,  with the logarithm of the surface gravity $\log(\rm g)=6$. 
 In Fig.~\ref{fsed}, we present a comparison between the SEDs  
 assumed in our models. The AGN SED maintains a
 high ionization flux even at high energies (low wavelengths) somewhat
 similar to the p-AGB one with the highest $T_{\rm eff}$ value. Some soft 
emission is noted for p-AGB stars with 100 kK and mainly with 50 kK. Both AGN and p-AGB SED models can be 
considered as the main ionizing source, i.e., responsible
for the ionization of the gas, and underlying stellar population was not considered in the models.
Therefore, our models are designed to investigate what kind of object would be producing
the gas ionization in UGC\,4805 based on emission line intensity ratios.
These models are not intended for analysing
the equivalent width of lines,  
as performed by \citet{2011MNRAS.413.1687C}, which also strongly depends on the underling  stellar population \citep{1981A&A...102..245D}.

 \item Metallicity: We assumed ($Z/{\rm Z_{\odot}}$) = 0.2, 0.5, 0.75, 1.0, 2.0, and 3.0 for the models.  We assumed the solar oxygen abundance to be 12 + $\log$ (O/H)$_\odot$ = 8.69 \citep{2009ARA&A..47..481A,Allende_Prieto_2001} and it is equivalent to ($Z/{\rm Z_{\odot}}$)=1.0. All the abundances of heavy elements were scaled linearly with the metallicity, except the nitrogen for which we assumed the relation $\rm \log(N/O)=1.29\times [12 + \log(O/H)] - 11.84$ derived by \citet{2017MNRAS.468L.113D}, who considered abundance estimations for type 2 AGNs and 
 \ion{H}{ii} regions. \\
 
\item Electron Density: We assumed for the models an electron density value of $N_{\rm e}$ = 500 $\rm cm^{-3}$, constant in the nebular radius. This value is very similar to the one estimated  for UGC\,4805 nucleus through the relation  between   $N_{\rm e}$ and   
$R_{S2}=$[\ion{S}{ii}]$\lambda 6716/\lambda 6731$ line ratio and using the {\sc IRAF/TEMDEN} task.
 Observational estimations of  $N_{\rm e}$ based
on the \ion{Ar}{iv}$\lambda$4711/$\lambda$4740 ratio, which map a denser gas region than the one
based on [\ion{S}{ii}] ratio, for two Seyfert nuclei (IC\,5063 and NGC\,7212)
by \citet{2017MNRAS.471..562C}, indicate  $N_{\rm e}$  ranging from $\sim 200$  to $\sim 13\,000 \: \, \rm cm^{-3}$.
Furthermore, radial gradients with electron densities decreasing from the centres to the edges have been
found in star-forming regions (e.g., \citealt{2000A&A...357..621C}) and in AGNs (e.g., \citealt{2018ApJ...867...88R}).
However, \citet{2020MNRAS.492.5675C} showed that models with $N_{\rm e} \: < 2\,000 \: \rm cm^{-3}$ produce practically the same optical emission-line ratios.  In addition, photoionization models  assuming electron density
variations along the radius have an almost negligible influence on predicted optical line
ratios as demonstrated by \citet{2019MNRAS.486.5853D}. For a detailed discussion about
the $N_{\rm e}$ influence on metallicity estimates in Seyfert 2 AGNs, see \citet{2020MNRAS.496.3209D}.

\item Ionization Parameter: This parameter is defined as 
\begin{equation}
\label{elogu}
U = \frac{Q({\rm H)}}{4 \, \pi \, R_{{\rm 0}}^2 \, n(\rm H) \, \rm  c},
\end{equation}
 in which $Q(\rm H)$ [$\rm s^{-1}$] is the number of hydrogen-ionizing photons emitted by the central object, $R_{0}$ [cm] is the distance from the ionization source to the inner surface of the ionized gas cloud, $n(\rm H)$ [cm$^{-3}$] is the total hydrogen density (ionized, neutral and molecular),  and $\rm c$ is the speed of light [cm\,s$^{-1}$]. We assumed logarithm of $U$ in the range of -4.0 $\le \log U \le $ -0.5, with a step of 0.5 dex, which is about the same range of values assumed
by \citet{2016MNRAS.456.3354F}, who used a photoionization model grid to reproduce ultraviolet and optical
emission-line ratios of active and normal galaxies. Different
ionization parameter values simulate  gas excitation differences, owing to variations in the mass of the gas phase and several geometrical conditions covering a wide range of possible scenarios 
\citep{2014MNRAS.441.2663P}. 

In our models, a plane-parallel geometry is adopted, and the outer radius is
assumed to be the one where the gas temperature falls to 4\,000 K
(default outer radius value in the CLOUDY code), since cooler gas practically does 
not produce optical emission lines. Models with different combinations of $Q(\rm H)$, $R_{0}$, and $n(\rm H)$, resulting in similar values of $U$, are homologous models, i.e., they  predict very similar emission-line intensities.

For the ionizing sources,  {\sc Cloudy}  is a unidimensional code that assumes a central ionization source, which is a good approach for AGNs. However, in giant star-forming regions (e.g., \citealt{2011MNRAS.413.2242M}),
stars are spreaded out through the gas. In this sense, in most cases, a central ionization source usage would not constitute a genuine representation of the situation.  \citet{2009Ap&SS.324..199E} and \citet{2008A&A...482..209J} showed
that the distribution  of the O-B stars in relation to the gas alters the ionisation structure and
the electron temperature. Hence, the ionization parameter partially depends on the distance of the ionizing source to the gas. However, in our case, we
are considering an integrated spectrum of the  UGC\,4805 nucleus; thus, the stellar distribution may have a minimum effect on the emergent spectrum.
In the case of giant  \ion{H}{ii} regions ionized by
  stellar clusters (e.g. \citealt{1996AJ....111.1252M, 2001A&A...380..137B}),
  the hottest stars dominate the gas ionization \citep{2017MNRAS.466..726D}.
Therefore, the assumption of a single star with a representative effective temperature as the main ionizing source, as assumed in our p-AGB models, is a good approximation \citep[see e.g.,][]{2019MNRAS.483.1901Z}.
\end{enumerate}

To estimate the O/H and $U$ values for the UGC\,4805 nucleus, we compare some observational emission line intensity ratios with 
photoionization model results using diagnostic diagrams and perform a linear interpolation between models.

\begin{figure*}
\centering
\includegraphics*[angle=0,width=.7\textwidth]{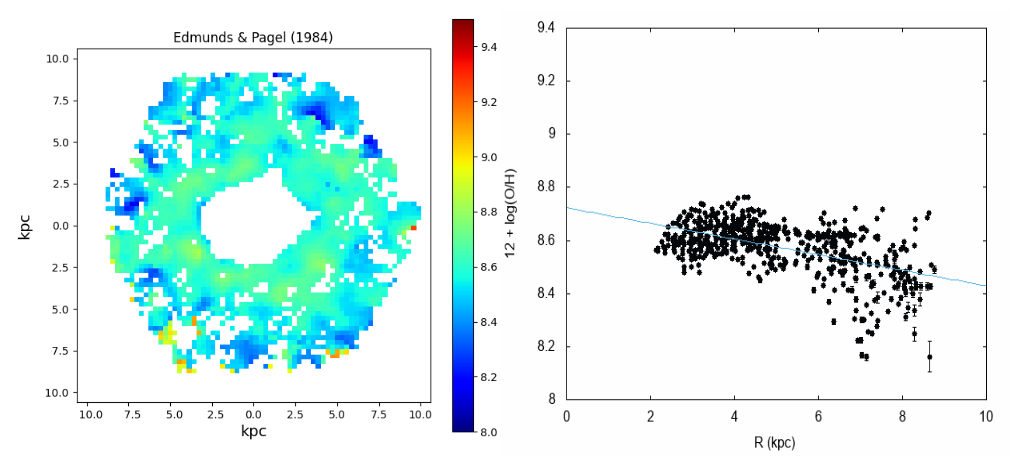}
\hspace*{-14pt}\includegraphics*[angle=0,width=.7\textwidth]{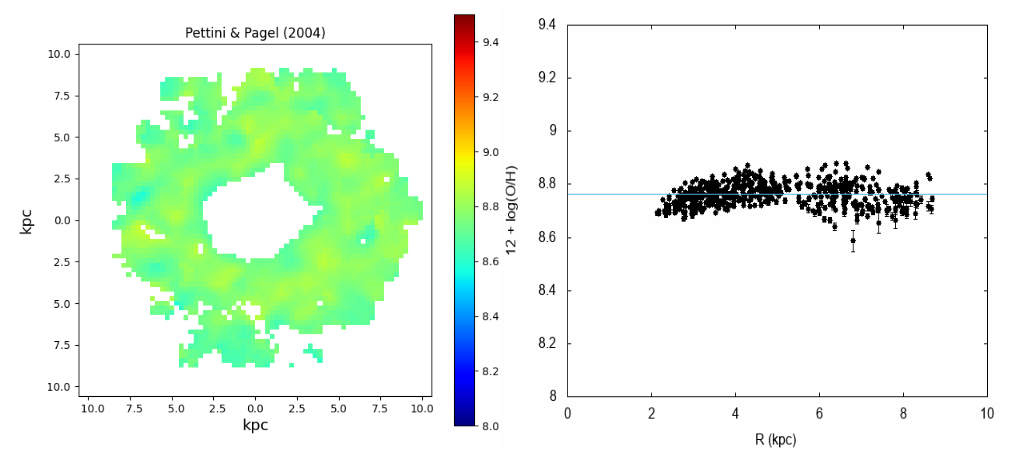}
\includegraphics*[angle=0,width=.7\textwidth]{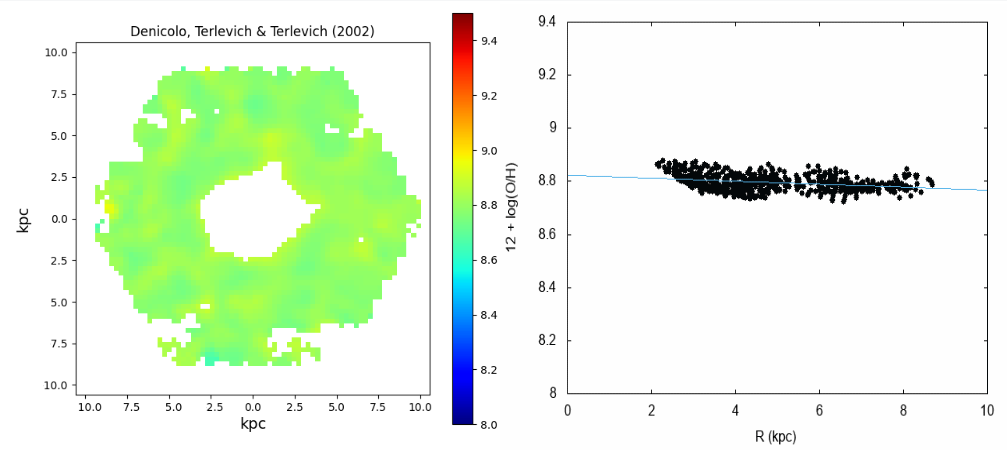}
\caption{Left panels: oxygen abundance maps obtained through the calibrations
described in Sect.~\ref{hiical} and  indicated in each plot.
Right panels: radial abundance distributions along the UGC\,4805 disk.
The line in each plot represents the linear fitting (Eq.~\ref{lingr}) to the estimations,  whose coefficients are listed in Table~\ref{resumo}.}
\label{extrapolacoes1}
\end{figure*}

\begin{figure*}
\centering
\includegraphics*[angle=0,width=.7\textwidth]{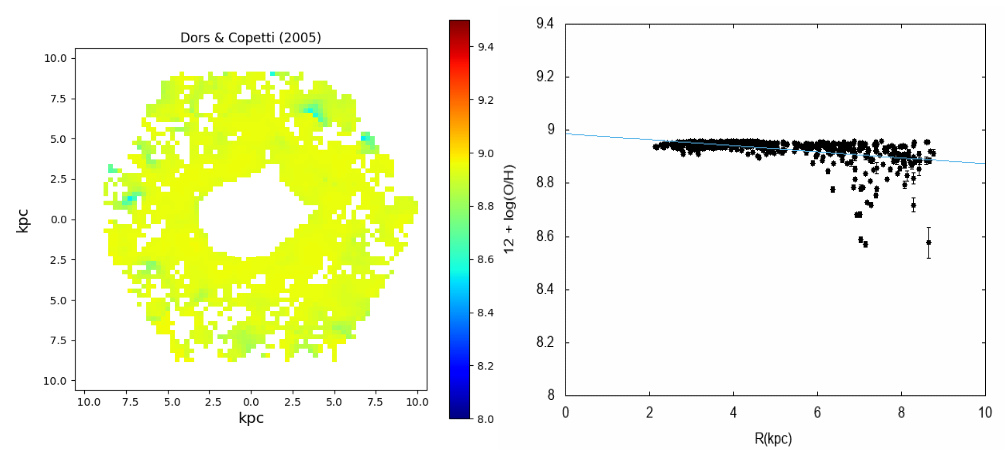}
\includegraphics*[angle=0,width=.7\textwidth]{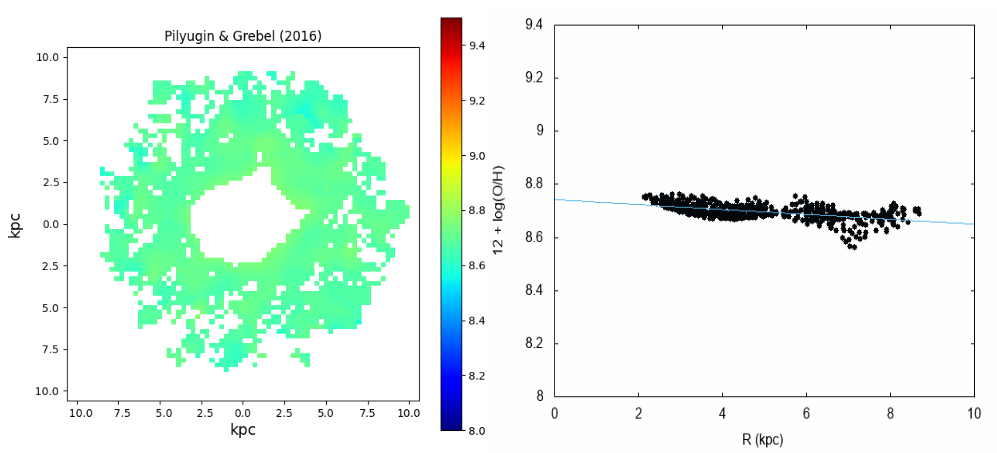}
\caption{Same as Fig.~\ref{extrapolacoes1}, but  for the indicated calibrations.}
\label{extrapolacoes2}
\end{figure*}

\begin{figure*}
\includegraphics*[angle=0,width=1\textwidth]{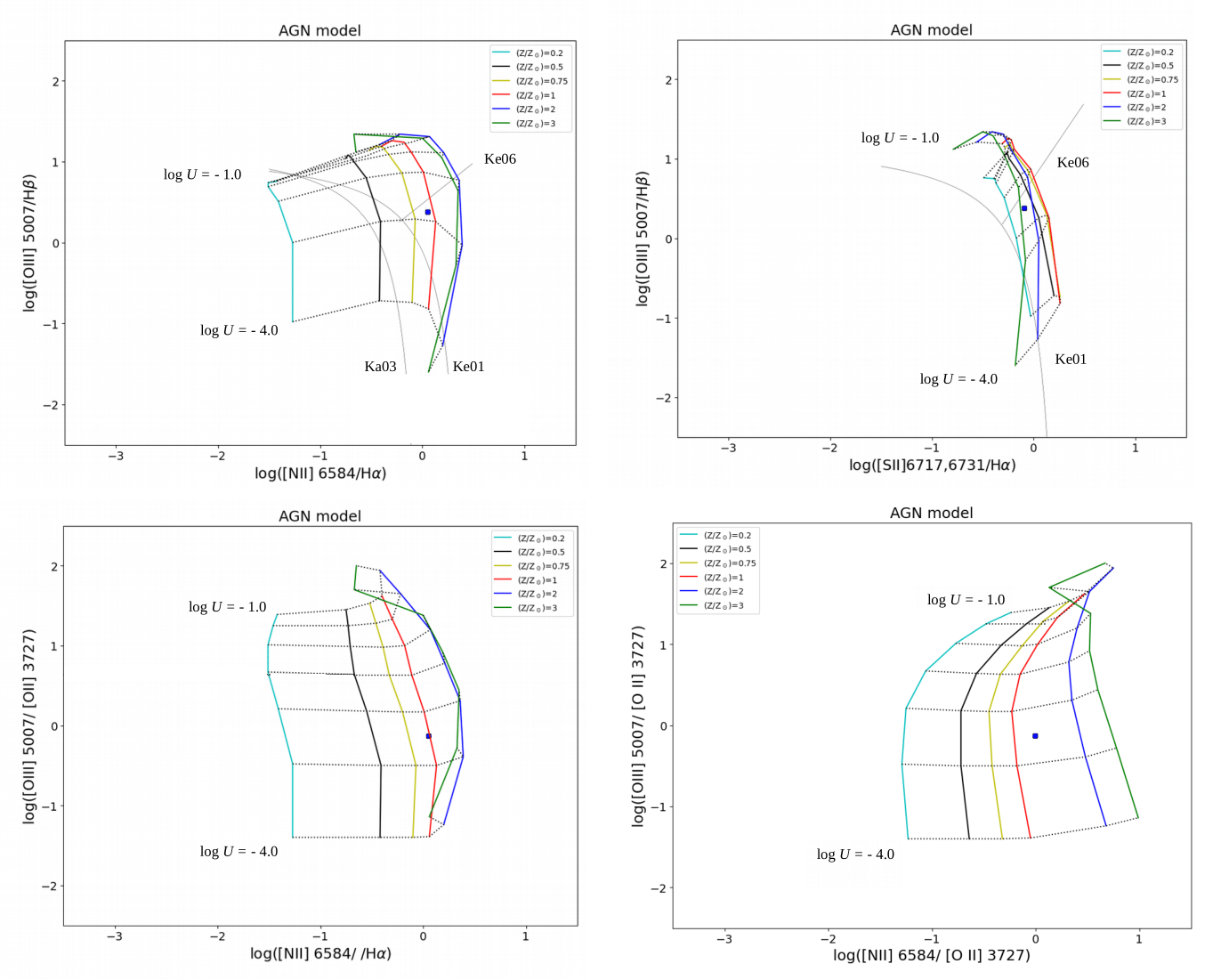}
\caption{Upper left panel: $\log$([\ion{O}{III}]$\lambda 5007$/H$\beta$) versus 
  $\log$([\ion{N}{II}]$\lambda 6584$/H$\alpha$) diagnostic diagram.
Upper right panel: $\log$([\ion{O}{III}]$\lambda 5007$/H$\beta$) versus 
$\log$([\ion{S}{II}]$\lambda\lambda 6717,6731$/H$\alpha$) diagnostic diagram. 
Gray lines  represent the separating criteria of the BPT diagrams, from \citet{2006MNRAS.372..961K} (Ke06), \citet{2003MNRAS.346.1055K} (Ka03), and  \citet{kewley01} (Ke01).  Lower left panel: $\log$([\ion{O}{III}]$\lambda 5007$/[\ion{O}{II}] $\lambda 3727$) versus $\log$([\ion{N}{II}]$\lambda 6584$/H$\alpha$) diagnostic diagram. Lower right panel:  $\log$([\ion{O}{III}]$\lambda 5007$/[\ion{O}{II}] $\lambda 3727$) versus 
$\log$([\ion{N}{II}]$\lambda 6584$/[\ion{O}{II}]$\lambda 3727$) diagnostic diagram.
Coloured solid lines connect AGN photoionization model results (see Sect.~\ref{modd}) with the same metallicity  $(Z/Z_\odot)$
and dotted lines  models with the same ionization parameter ($U$), as indicated. The blue point represents the observational line ratios for the UGC\,4805 nucleus (see Sect.~\ref{dataobs}).}
\label{modelos}
\end{figure*}


\begin{figure*}
\includegraphics*[angle=0,width=1\textwidth]{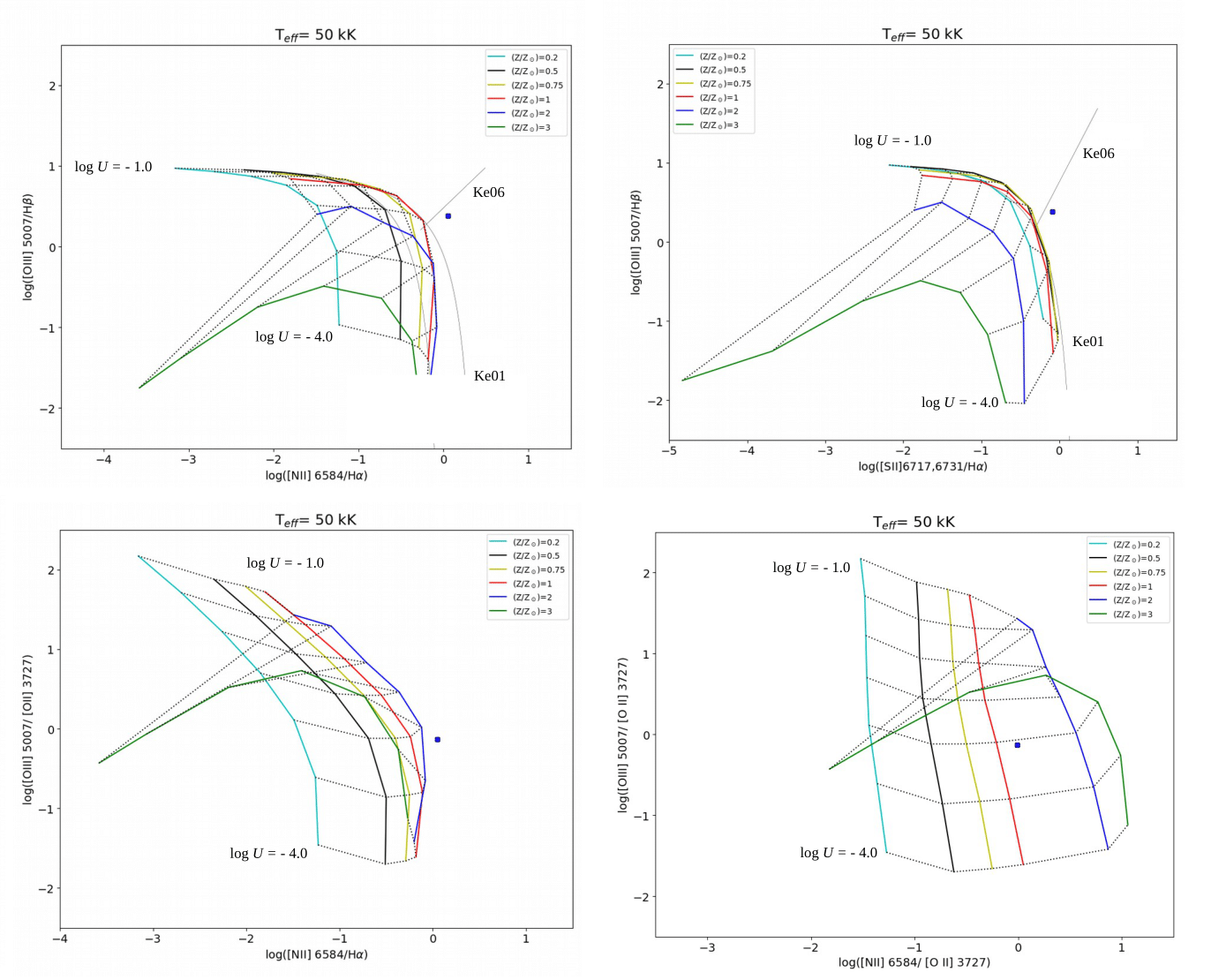}
\caption{Same as Fig.~\ref{modelos} but considering p-AGB photoionization models (see Sect.~\ref{modd}) assuming  $T_{\rm eff}$ = 50 kK.}
\label{modelos_50k}
\end{figure*}

\begin{figure*}
\includegraphics*[angle=0,width=1\textwidth]{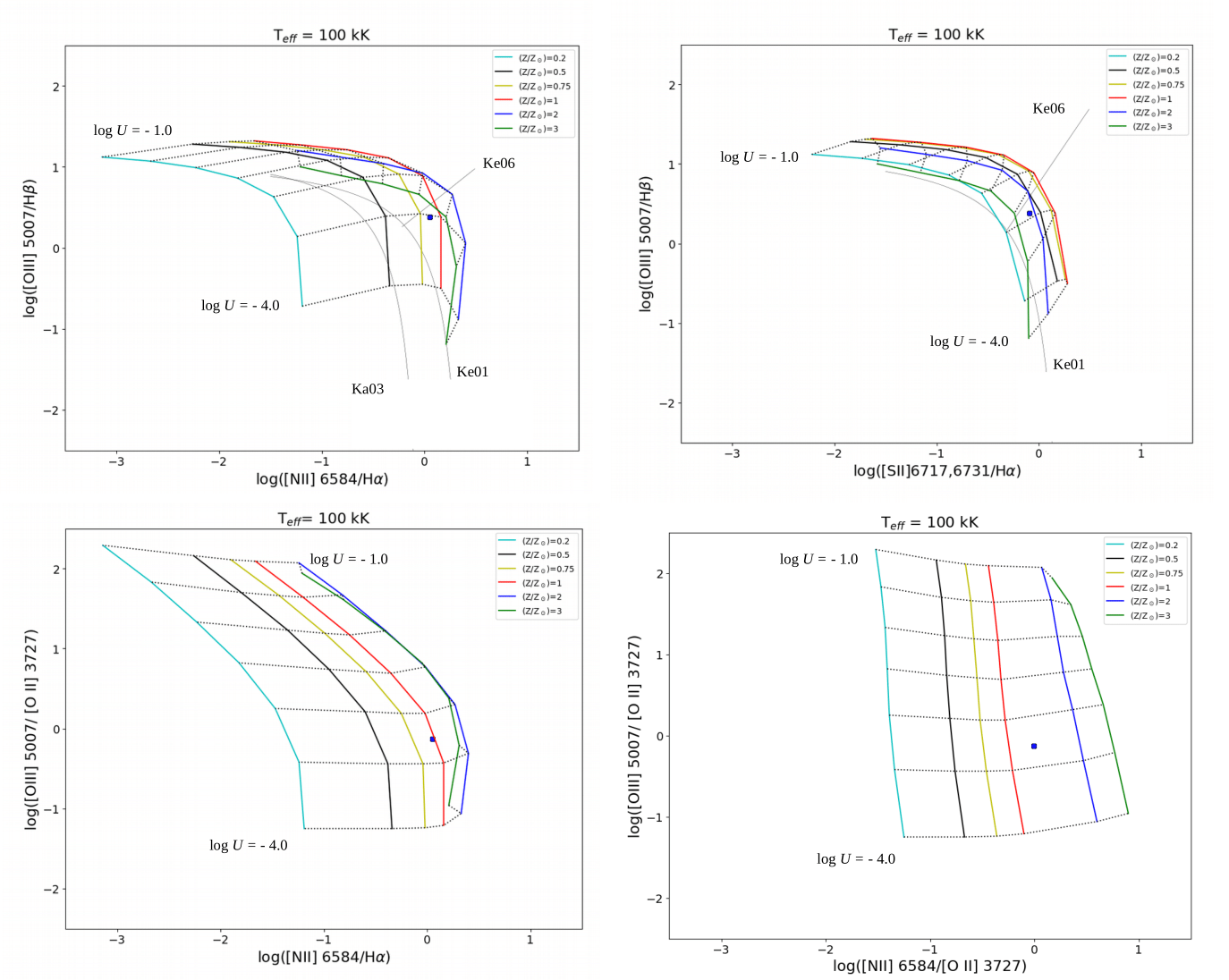}
\caption{Same as Fig.~\ref{modelos_50k} but considering p-AGB photoioniazation models (see Sect.~\ref{modd}) assuming  $T_{\rm eff}$ = 100 kK.}
\label{modelos_100k}
\end{figure*}

\begin{figure*}
\includegraphics*[angle=0,width=1\textwidth]{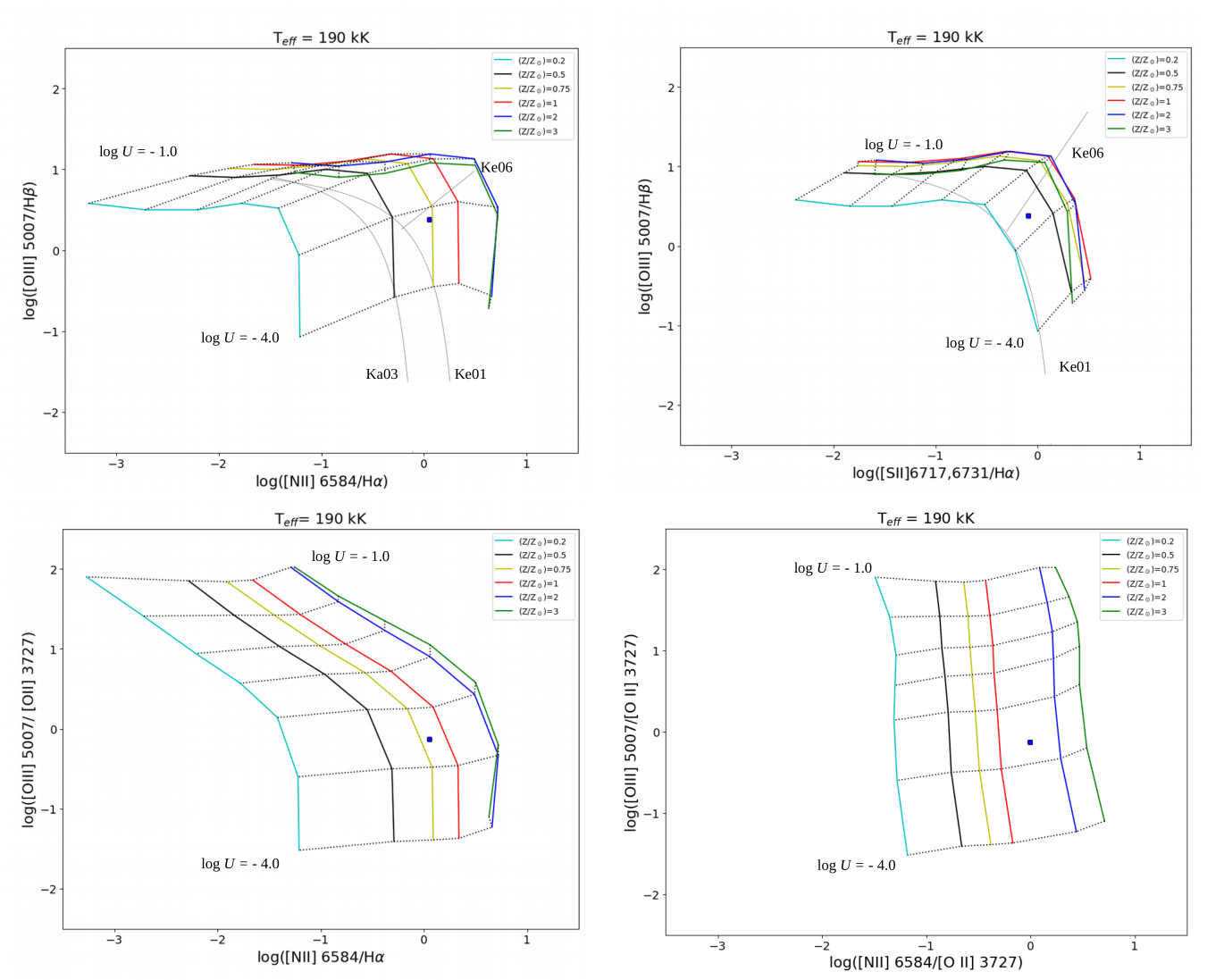}
\caption{Same as Fig.~\ref{modelos_50k} but considering p-AGB photoioniazation models (see Sect.~\ref{modd}) assuming  $T_{\rm eff}$ = 190 kK.}
\label{modelos_190k}
\end{figure*}

%
%
%

\section{Results}
\label{res}

\subsection{O/H calibrations}

To apply some of the strong-line calibrations developed for SFs described in Sect.~\ref{hiical} to the UGC\,4805 disk \ion{H}{ii} regions, it is necessary to select
which branch of the (O/H)-$R_{23}$ relation is adequate. We consider the \cite{Kewley_2008} criteria to break the degeneracy, i.e., for objects with 
 $\log$([\ion{N}{ii}]$\lambda$6584/[\ion{O}{ii}]$\lambda$3727)$\: > \:-1.2$, the  upper $R_{23}$ branch must be used. 
  The O/H abundances were estimated only for objects classified as pure star-forming regions, i.e., those with line ratios under the \citet{2003MNRAS.346.1055K} line in the diagnostic diagram in the left panel of 
 Fig.\ \ref{bpt_diag}. Figures.~\ref{extrapolacoes1} and ~\ref{extrapolacoes2} present the abundance maps (left panels) and the O/H values along the disk (right panels).
Note that all the strong-line calibrations applied exhibited a 
  linear decrease of O/H along the disk in agreement with previous results  (e.g., \citealt{2004A&A...425..849P}). We derive the central oxygen abundance 
  $\rm 12 + \log(O/H)_{0}$ extrapolating to the centre of the galaxy the linear fit:

\begin{equation}
\label{lingr}
{\rm  12 + \log (O/H)=12 +\log(O/H)_{0}}+ (grad \: \times R),
\end{equation}
where $\rm 12 + \log (O/H)$ is the oxygen abundance at a given
galactocentric  distance $R$ (in units of arcsec) and $grad$ is the regression slope. 
The parameters of the linear regressions for the distinct calibrations used are listed in Table~\ref{resumo}. The star-forming ring, clearly visible
in the H$\alpha$ map (see Fig.~\ref{figure1}), does not present any oxygen abundance discrepancy in comparison to its neighbour regions.

The calibration proposed by \cite{edmunds84} resulted in 12 + $\log$(O/H) values ranging from $\rm \sim 8.2$ to $\sim 8.8$ along the galactic disk, while the abundance value extrapolated to the nucleus  ($R=0$ arcsec) is $\rm  12 + \log(O/H)_{0} = 8.72 $.
Considering the  \cite{denicolo} calibration, we derive 
12 + $\log$ (O/H)$_{0}$ = 8.81  for the nucleus.  
By using the calibration by \cite{2004MNRAS.348L..59P}, we derive
a nuclear abundance of 12 + $\log$ (O/H)$_{0}$ = 8.79.
Estimates of oxygen abundances obtained using the calibration by  \cite{dors} yield a flatter gradient than the gradients derived with other calibrations, i.e., O/H values vary along the galactic disk in the narrow range of 8.85 < 12 + $\log$(O/H) < 9.0. 
The estimated nuclear abundance is 12 + $\log$ (O/H)$_{0}$ = 8.98. Note that a large part of our estimated values are close to the upper metallicity limit for this calibration, where the metallicity is practically constant, i.e., the O/H abundance is saturated with the variation of 
$R_{23}$. Finally, the application of the calibration proposed by \cite{pilyugin16} indicates abundances in the range of  8.5 < 12 + $\log$(O/H) < 8.7, with an inferred central abundance of 12 + $\log$ (O/H)$_{0}$ = 8.76, which is close to the abundance obtained through the  \cite{2004MNRAS.348L..59P} calibration.  In summary, the extrapolation for the UGC\,4805 LINER nucleus based on  the calibrations considered above indicates an over solar oxygen abundance, with an averaged value of $\rm 12+\log(O/H)_{0}=8.82$.

To estimate the O/H abundance by using the nuclear emission of UGC\,4805, we used the line intensity ratios listed 
in Table~\ref{fluxos} and applied the  \cite{Storchi_Bergmann_1998}, \cite{10.1093/mnras/stx150}, and
\citet{2020MNRAS.492.5675C} calibrations. The estimated values of O/H abundance are  listed in Table~\ref{resumo}.
As suggested by \cite{Storchi_Bergmann_1998}, the final (O/H) abundance derived from their methods should be the average of the values calculated from the two equations (Sect.~\ref{agnc}), which provides  12 + $\log$ (O/H)$_{0}$ $= 8.93  \pm$ 0.04.
The \cite{10.1093/mnras/stx150} and \citet{2020MNRAS.492.5675C} calibrations provide a value of 12 + $\log$ (O/H)$_{0}$ = $8.77 \pm$ 0.01 and  
12 + $\log$ (O/H)$_{0}$ = $8.69 \pm$ 0.01, respectively.  
An average value of  12 + $\log$ (O/H)$_{0}=8.81 \pm 0.02$ was derived considering the three calibrations.

\begin{table*}
\centering
\caption{Oxygen abundance results for the UGC\,4805 nucleus.
The first set of values are the central Z/Z$_{\odot}$ estimations and the coefficients of the linear fitting (Eq.~\ref{lingr})
 to the O/H estimations along the UGC\,4805 disk (see Figs.~\ref{extrapolacoes1} and \ref{extrapolacoes2})
 considering different calibrations for \ion{H}{ii} regions proposed by different authors as indicated (see Sect.~\ref{hiical}).
The second set of values are the metallicities, the oxygen abundances, and $\log U$ (only for one case) obtained by using the AGN calibrations (see Sect.~\ref{agnc}). The third set of values are metallicities, O/H abundances and $\log U$ obtained from linear interpolations of the  photoionization model results shown in Fig.~\ref{modelos}, \ref{modelos_50k},
\ref{modelos_100k} and \ref{modelos_190k}. The diagnostic diagrams and the model ionizing sources considered are indicated.}
\label{resumo}
\begin{tabular}{@{}llll@{}}
\noalign{\smallskip}
\hline
\multicolumn{4}{c}{Central intersect method -- \ion{H}{ii} region calibrations}    \\
& \multicolumn{1}{c}{Z/Z$_{\odot}$} & \multicolumn{1}{c}{12 + $\log$(O/H)$_{0}$}  & \multicolumn{1}{c@{}}{\textit{grad} (dex/arcsec)} \\ 
\cite{edmunds84}  &1.07& 8.72 $\pm$ 0.003  & $-0.016 \pm 0.001$   \\
\cite{denicolo} &1.32& 8.81 $\pm$ 0.002  & $-0.002 \pm 0.0001$    \\
\cite{2004MNRAS.348L..59P} &1.26& 8.79 $\pm$ 0.003 & $-0.003 \pm 0.0002$   \\
\cite{dors} &1.95& 8.98 $\pm$ 0.002 & $-0.004 \pm 0.0002$   \\
\cite{pilyugin16} &1.17& 8.76 $\pm$ 0.002 & $-0.007 \pm 0.0001$   \\[2pt]
Average & 1.35 & 8.82 $\pm$ 0.003& \\  
\hline
\multicolumn{4}{c}{AGN calibrations}    \\
                              & Z/Z$_{\odot}$ & 12 + $\log$(O/H) & $\log U$\\
\cite{Storchi_Bergmann_1998} &1.74& $8.93$ $\pm$ 0.04 & ---  \\
\cite{10.1093/mnras/stx150}  &1.20& $8.77 \pm 0.01$ & --- \\
\citet{2020MNRAS.492.5675C}  &1.00& $8.69 \pm 0.01$ &  $-3.09$ \\[2pt]
Average &1.31&8.81 $\pm$ 0.02& \\  
\hline
\multicolumn{4}{c}{Diagnostic diagrams -- Photoionization models}  \\
                                & Z/Z$_{\odot}$ & 12 + $\log$(O/H)& $\log U$ \\
\multicolumn{4}{l}{AGN models}  \\
$\log([\ion{O}{iii}]/\rm H\beta)$ vs.\ $\log$([\ion{N}{ii}]/H$\alpha$) &0.95& $8.67 \pm 0.02$ & $-3.39$  \\
 
$\log([\ion{O}{III}]/[\ion{O}{II}]) $ vs.\ $\log$($[\ion{N}{II}]$/H$\alpha$) &0.93& $8.66  \pm 0.02$ & $-3.22 $  \\

$\log([\ion{O}{III}]/[\ion{O}{II}])$ vs.\ $\log([\ion{N}{II}]/[\ion{O}{II}])$ &1.29& $8.80  \pm 0.02$& $-3.24$  \\[2pt]
  
Average &1.06 & $8.71  \pm 0.02$ & \\[4pt]
%
\multicolumn{4}{l}{p-AGB models ($T_{\rm eff}$= 100 kK)}  \\
$\log([\ion{O}{iii}]/\rm H\beta)$ vs.\ $\log$([\ion{N}{ii}]/H$\alpha$) &0.85& $8.62  \pm 0.03$ & $-3.50$ \\

$\log([\ion{O}{III}]/[\ion{O}{II}])$ vs.\ $\log$($[\ion{N}{II}]$/H$\alpha$) &0.98& $8.68  \pm 0.01$ & $-3.26 $  \\
$\log([\ion{O}{III}]/[\ion{O}{II}])$ vs.\ $\log([\ion{N}{II}]/[\ion{O}{II}])$ &1.32& $8.81  \pm 0.02$ & $-3.29$ \\[2pt]
Average &1.06& $8.71  \pm 0.02$ & \\[4pt]
\multicolumn{4}{l}{p-AGB models ($T_{\rm eff}$= 190 kK)}  \\
$\log([\ion{O}{iii}]/\rm H\beta)$ vs.\ $\log$([\ion{N}{ii}]/H$\alpha$) &0.72& $8.55  \pm 0.01$  & $-3.57$ \\
$\log([\ion{O}{III}]/[\ion{O}{II}])$ vs.\ $\log$($[\ion{N}{II}]$/H$\alpha$) &0.81& $8.60  \pm 0.01$ & $-3.26 $ \\
$\log([\ion{O}{III}]/[\ion{O}{II}])$ vs.\ $\log([\ion{N}{II}]/[\ion{O}{II}])$ &1.48& $8.86  \pm 0.01$ & $-3.31$ \\[2pt]
Average &1.00 &$8.69 \pm 0.01$ & \\[4pt]
\hline
\end{tabular}
\end{table*}


\subsection{Photoionization models}

As mentioned previously (see Sect.~\ref{modd}), two different photoionization model grids were built, 
one assuming an AGN as the ionizing source and another assuming p-AGB stars with different $T_{\rm {eff}}$ values as the ionizing source. 
In the upper panels of Fig.~\ref{modelos},  the observational line ratios of the
UGC\,4805 nucleus are plotted in the $\log$([\ion{O}{III}]$\lambda 5007$/H$\beta$) versus $\log$([\ion{N}{II}]$\lambda 6584$/H$\alpha$) (left panel) and $\log$([\ion{O}{III}]$\lambda 5007$/H$\beta$) versus $\log$([\ion{S}{II}]$\lambda \lambda 6717,6731$/H$\alpha$) (right panel) diagnostic diagrams and compared to those predicted by AGN photoionization models.  These plots also show the demarcation lines proposed by  \citet{2003MNRAS.346.1055K} and \citet{2006MNRAS.372..961K}. The observational line intensity ratios are reproduced by the AGN models; therefore, we can infer 
a metallicity and an ionization parameter for the UGC\,4805 nucleus. Using linear interpolation between the models in the $\log$([\ion{O}{III}]/H$\beta$) versus $\log$([\ion{N}{II}]/H$\alpha$) diagnostic diagram (Fig.~\ref{modelos} upper left panel), we derive a metallicity of  \textit{(Z/Z$_\odot$) $\sim$ 0.95}  and  $\log U \sim -3.39$.  For 
$\log$([\ion{O}{III}]/H$\beta$) versus $\log$([\ion{S}{II}]/H$\alpha$) diagnostic diagram  (Fig.~\ref{modelos} upper right panel), which is clearly bi-valuated with the upper envelope at    (Z/{$\rm Z_\odot) \sim$ 1},
we adopt the models with larger values to characterise our object,  since the low metallicity models do not represent AGN-like objects, as it is seen in the left panel. Then, we derived  \textit{(Z/Z$_\odot$)} $\sim$ 2.57 and $\log U \sim -3.26$, using the $\log$([\ion{O}{III}]/H$\beta$) versus $\log$([\ion{S}{II}]/H$\alpha$) diagnostic diagram.  The second metallicity value is about three times the former one.

\citet{2011MNRAS.415.3616D}, by using a grid of photoionization models, showed that there are relations between different line ratios, such as $[\ion{O}{III}]\lambda 5007$/$[\ion{O}{II}] \lambda 3727$ versus $[\ion{N}{II}]\lambda 6584$/H$\alpha$ and 
 [\ion{O}{III}]$\lambda 5007$ / [\ion{O}{II}] $\lambda 3727$ versus [\ion{N}{II}]$\lambda 6584$/ [\ion{O}{II}]$\lambda 3727$,
that are more sensitive to the ionization parameter, and the metallicities obtained through them are closer to those obtained using the $T_{\rm e}$-method.
For this reason, we use these diagnostic diagrams also employed  by \citet{10.1093/mnras/stx150} and \citet{2020MNRAS.492.5675C} to perform a more reliable analysis. The lower panels of Fig.~\ref{modelos} presents these observational line ratios for the UGC\,4805 nucleus superimposed on those ratios predicted by our AGN photoionization models.
By using linear interpolation between the models 
  we derive   \textit{(Z/{\rm Z}$_\odot$) $\sim$ 0.93} and $\log U \sim -3.22$ from the $\log$([\ion{O}{III}]/[\ion{O}{II}]) vs.\ $\log$([\ion{N}{II}]/H$\alpha$) diagnostic diagram (lower left panel), and  \textit{(Z/{\rm Z}$_\odot$) $\sim$ 1.29} and  $\log U \sim -3.24$ from the $\log$([\ion{O}{III}]/[\ion{O}{II}]) vs.\ $\log$([\ion{N}{II}]/[\ion{O}{II}]) diagnostic diagram (lower right panel).

The values of the ionization parameter found using the four diagnostic diagrams (Fig.\ \ref{modelos}) are very similar and in agreement with the typical value for LINER galaxies estimated by \cite{1983ApJ...264..105F}.

Figs.~\ref{modelos_50k}, \ref{modelos_100k},  and \ref{modelos_190k} contain the same diagnostic diagrams exhibited in Fig.~\ref{modelos} for the photoionization model results considering  p-AGB stars as ionizing sources.
In Fig.~\ref{modelos_50k}, the models with $T_{\rm eff}= 50$ kK do not reproduce the UGC\,4805 nucleus line ratios. In the upper panels of this figure, the parameter space characterized by the models is  occupied only by \ion{H}{ii}-like objects. Therefore, it is impossible to derive any value
of $Z$ or $U$ from these models.
For models with $T_{\rm eff}= 100$ and 190 kK
  and considering the  $\log$([\ion{O}{III}]/H$\beta$) versus $\log$([\ion{N}{II}]/H$\alpha$) (upper left panels of Figs.\ \ref{modelos_100k} and \ref{modelos_190k}), we derive  \textit{(Z/}Z$_{\odot}$) $\sim$ 0.85 and $\log U\sim 3.50$, and  \textit{(Z/}Z$_{\odot}$) $\sim$ 0.72  and $\log U\sim 3.57$, respectively.
Taking into account $T_{\rm eff}= 100$ kK  and  $\log$([\ion{O}{III}]/H$\beta$) versus $\log$([\ion{S}{II}]/H$\alpha$) diagnostic diagram  (upper right panel),  we found $\log U \sim -3.44$ and two values for the metallicity, i.e.,   Z/$\rm Z_{\odot}\sim$ 2.87
and  Z/$\rm Z_{\odot}\sim$ 0.42.
This happens because, as in the case of AGN models, this relation is bi-valuated for the metallicity.
Analysing the results of the same diagnostic diagram for the p-AGB models with $T_{\rm eff}= 190$ kK, we do not observe a bi-valuated relatio. Models with metallicities larger than 0.75 occupy almost the same region. We obtain  \textit{(Z/Z$_\odot)$} $\sim$ 0.36 and $\log U=-3.35$.
These results could indicate that the high metallicity model solution found for the models with T$_{\rm eff}= 100$ kK [$(Z/\rm Z_{\odot})\sim 2.0$] is not correct.

The lower panels of Figs.~\ref{modelos_100k} and \ref{modelos_190k} display the same diagnostic diagrams as in the lower panels of Fig.~\ref{modelos}, but containing photoionization model results considering p-AGB stars as ionizing source.
For models with  $T_{\rm eff}= 100$ kK (Fig.\ \ref{modelos_100k}), we derive from the $\log$([\ion{O}{III}]/[\ion{O}{II}]) versus 
$\log$([\ion{N}{II}]/H$\alpha$) diagnostic diagram  Z/$\rm Z_{\odot} \sim$ 0.98 and  $\log U \sim -3.26$. From the $\log$([\ion{O}{III}]/[\ion{O}{II}]) versus 
$\log$([\ion{N}{II}]/[\ion{O}{II}]) diagram we calcule  Z/$\rm Z_{\odot}\sim$ 1.32 and  $\log U \sim -3.29$. Finally, we see that the models with $T_{\rm eff}= 190$ kK (Fig.\ \ref{modelos_190k}) provide from the $\log$([\ion{O}{III}]/[\ion{O}{II}]) versus
$\log$([\ion{N}{II}]/H$\alpha$) diagram a metallicity of  Z/$\rm Z_{\odot}\sim$ 0.81 and  $\log U \sim -3.26$, and from the $\log$([\ion{O}{III}]/[\ion{O}{II}]) versus 
$\log$([\ion{N}{II}]/[\ion{O}{II}]) diagram   Z/$\rm Z_{\odot} \sim$ 1.48 and  $\log U \sim -3.31$. 

The models yield bi-valuated or saturated results for the 
emission-line diagnostic diagrams that include the [\ion{S}{ii}] emission-lines and  show the more discrepant results including super-solar metallicities values [$(Z/Z_\odot) \sim$ 2.57] for the AGN models and sub-solar metallicities for the p-AGB models with  $T_{\rm eff}= 100$ and 190 kK  [$(Z/Z_\odot) \sim $ 0.42, 0.36, respectively]. Hence, we do not take into account the results derived from the $\log$([\ion{O}{III}]/H$\beta$) versus $\log$([\ion{S}{II}]/H$\alpha$) diagnostic diagrams.
The adopted $(Z/Z_\odot)$, 12 + log(O/H) and $\log U$ values derived from Figs.~\ref{modelos}, \ref{modelos_100k}, and \ref{modelos_190k} are listed in Table~\ref{resumo}.

The averaged values obtained from the extrapolation of the oxygen abundance gradient from \ion{H}{ii} region estimations and from AGN calibrations are   Z/Z$_\odot) \sim$ 1.35 and   (Z/Z$_\odot) \sim$ 1.31, respectively. 
In both cases, the estimated abundance values are over-solar and are in agreement, taking into account their errors.
On the other hand, all the photoionization model produce very similar average values close to the solar one:  \textit{(Z/Z$_\odot)$ $\sim$ 1.06, 1.06, 1.00} for AGN, p-AGB with $T_{\rm eff}= 100$ kK, and p-AGB with $T_{\rm eff}= 190$ kK, respectively.

\section{Discussion}
\label{disc}

A widely accepted practice is to estimate the oxygen abundance at the central part of a galaxy by the central intersect abundance [$\rm 12 + \log(O/H)_{0}$] obtained from the radial abundance gradient 
(e.g., \citealt{1992MNRAS.259..121V, 1994ApJ...420...87Z, 1998AJ....116.2805V}).  This methodology has predicted  solar or slightly over-solar metallicities for the central region of spiral galaxies, i.e.,  12 + $\log$(O/H)   from $\sim8.6$ to $\sim9.0$ (e.g., \citealt{2004A&A...425..849P, 2020MNRAS.492..468D}), depending on the method considered to derive the individual disk \ion{H}{ii}-region abundances. Comparisons of these extrapolated oxygen abundance measurements 
($\rm 12 + \log(O/H)_{0}$) with the ones obtained through the use of other methods that directly involve the nuclear emission have achieved good agreement. 
\cite{Storchi_Bergmann_1998} found that the O/H abundances derived for a sample
of seven Seyfert 2 galaxies through their calibrations are in consonance with those obtained 
by the central intersect abundance. This agreement was also found by 
\citet{2015MNRAS.453.4102D} using a larger sample of objects than the one considered
by \cite{Storchi_Bergmann_1998}.

The oxygen abundance profile along the UGC\,4805 disk presents a negative gradient, as expected, since it is a spiral galaxy. The negative gradient is explained naturally by models assuming the inside-out scenario of galaxy formation \citep{portinari,macarthur,barden}. According to this scenario, galaxies begin to form in the inner regions before the outer regions. This was confirmed by studies of the stellar populations  (e.g., \citealt{boissier, bell, pohlen}) and chemical abundances of spiral galaxies
(e.g., \citealt{2014A&A...563A..49S}). As previously shown, considering the O/H gradient extrapolation, AGN calibrations, and AGN and p-AGB photoionization models, we derived averaged oxygen abundance values for the UGC\,4805 nucleus in the range of   1.00 $\: < \:$ (Z/Z$_{\odot}) \: <$ 1.35, i.e., ranging from solar to slightly over-solar metallicities.  

\begin{figure}
\includegraphics*[width=0.4\textwidth]{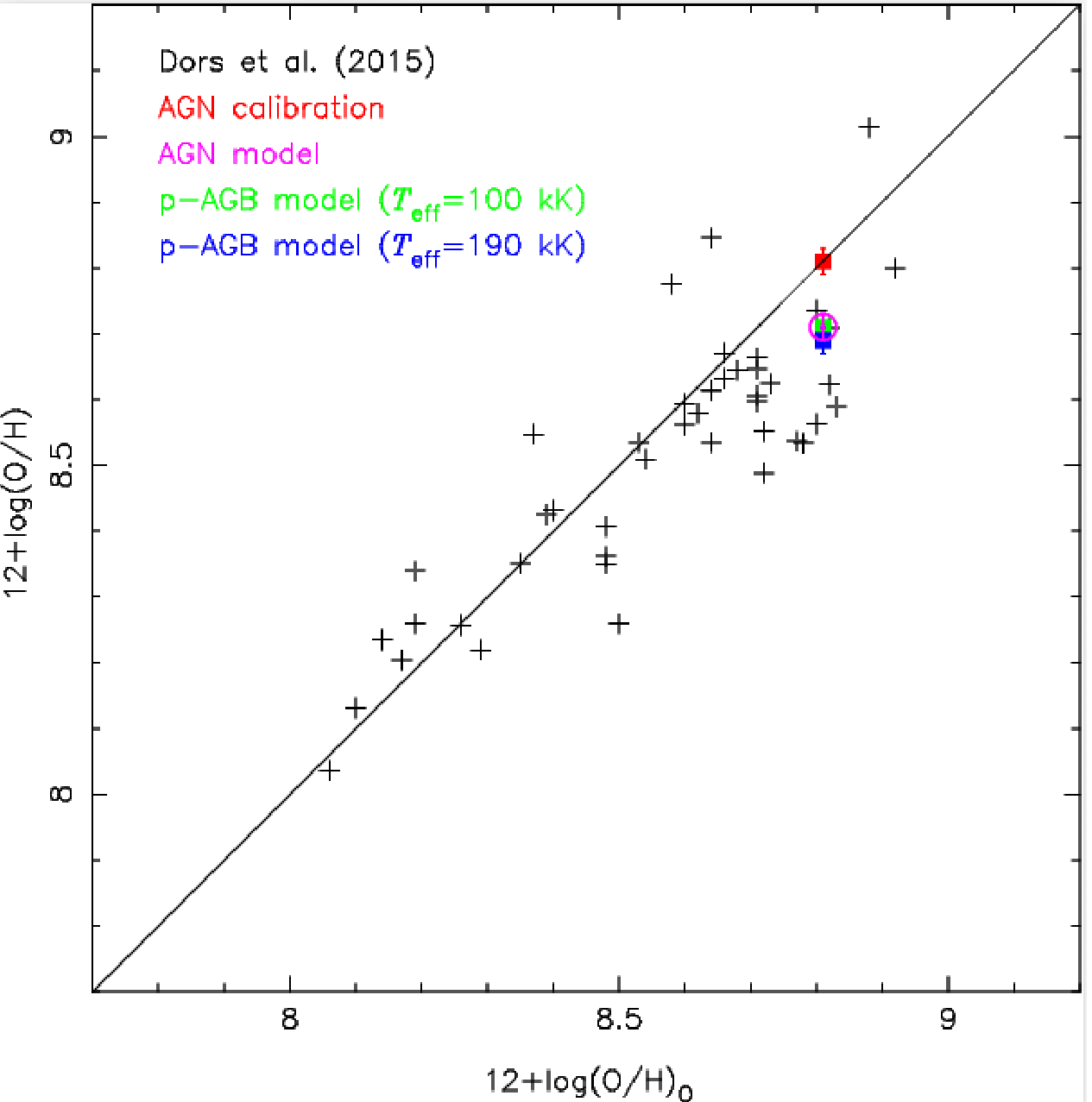}
\caption{Comparison between central intersect oxygen abundances derived for the UGC\,4805 nucleus from the radial abundance gradients ($12 + \log(\rm O/H)_{0}$) with those derived through
strong-line methods and AGN and p-AGB models (colored points as indicated). The point of the AGN model is the average from the AGN models.
Black points represent the estimations performed by
\citet{2015MNRAS.453.4102D} using the observational data by \citet{1997ApJS..112..315H}.
Solid line represents the equality between the estimations.}
\label{figco}
\end{figure}

\begin{figure}
\includegraphics*[angle=-90,width=0.4\textwidth]{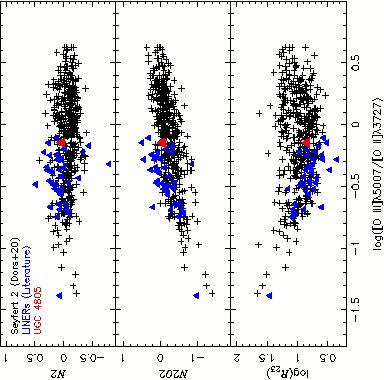}
\caption{Metallicity sensitive line ratios $R_{23}$, $N2O2$ and $N2$ versus the ionization parameter sensitive line ratio [\ion{O}{iii}]$\lambda$5007/[\ion{O}{ii}]$\lambda$3727. Black points represent 463 Seyfert 2 nuclei
studied by \citet{2020MNRAS.492..468D} and blue points represent 38 LINERs compiled
by \citet{1993ApJ...417...63H}, \citet{2001ApJ...554..240E}, \citet{2010A&A...519A..40A}, 
and \citet{2018ApJ...864...90M}. The red point represents the UGC\,4805 nucleus.}
\label{figsl}
\end{figure}

In Fig.~\ref{figco} the O/H average values estimated for the UGC\,4805 nucleus using AGN calibrations as well as  AGN and p-AGB models are compared with the average value derived through the central intersect method.  
The estimations for active and star-forming nuclei from \citet{2015MNRAS.453.4102D} are also presented in Fig.~\ref{figco}. This figure clearly illustrates that the averaged O/H value derived through the central intersect method is in consonance with the ones derived through the use of  AGN calibrations and AGN and p-AGB models, as 
well as with the \citet{2015MNRAS.453.4102D} estimations.

\citet{2010A&A...519A..40A} compared  intermediate-resolution optical spectra of a sample of 49 nuclei classified
as LINERs/composites with photoionization
model results assuming as ionization source accretion-rate AGN (represented by a power law SED) using the \citet{2004ApJS..153...75G} models and the shock models built by \citet{2008ApJS..178...20A}.
These authors also compared the observed and predicted equivalent widths of the lines present on their   spectra using models with p-AGB SEDs computed by \citet{1994A&A...292...13B} [see also \citealt{2009ASPC..408..122C}], finding that photoionization by p-AGB stars alone can explain
only  $\approx 22$\% of the observed LINER/composite sample. They also found that the major fraction of their sample could be characterized by nuclear emission consistent with excitation by a low-accretion rate AGNs and/or fast shocks.
\citet{2018ApJ...864...90M} compared observational  optical and ultraviolet spectra of three
LINERs with model results assuming four different excitation
mechanisms: shocks, photoionization by an accreting black hole, and photoionization by young or old hot stars. These authors concluded that the model which  best describes their data has a low-luminosity accretion-powered active nucleus that photoionizes the gas within $\sim 20$ pc of the galaxy centre, as well as shock excitation of the gas at larger distances.
These authors also indicated that LINERs could have more than one ionizing mechanism. In the case of the UGC\,4805 nucleus, the good agreement among all the different methods applied to derive its metallicity does not allow discrimination of the nature of the ionizing source.

Fig.~\ref{figsl} illustrates  the $\log(R_{23})$, $N2O2$ and $N2$ metallicity indexes as a function of the
[\ion{O}{iii}]$\lambda$5007/[\ion{O}{ii}]$\lambda$3727 line ratio used as an ionization 
parameter indicator  for the UGC\,4805 nucleus. 
This figure compares
our results to those of a sample of confirmed 463 Seyfert 2 nuclei studied by \citet{2020MNRAS.492..468D} and obtained from the Sloan Digital Sky Survey \citep{2000AJ....120.1579Y}, as well as  those of a sample 
 of 38 LINERs obtained by \citet{1993ApJ...417...63H}, \citet{2001ApJ...554..240E}, \citet{2010A&A...519A..40A}, and \citet{2018ApJ...864...90M}.
Both populations LINERs and Seyfert 2s, are partially overlapped in all of these diagrams although they display slightly different trends with LINERs showing lower ionizations ($\log U \: < \: -3.2$) following  Eq.\ \ref{eq2}. 
As can be seen  in Fig.~\ref{figsl}, the UGC\,4805 nucleus positions in these diagrams are compatible with both populations, although they seems to follow the LINERs sequence; therefore, they would share similar physical properties.

\begin{figure}
\includegraphics*[angle=-90,width=0.4\textwidth]{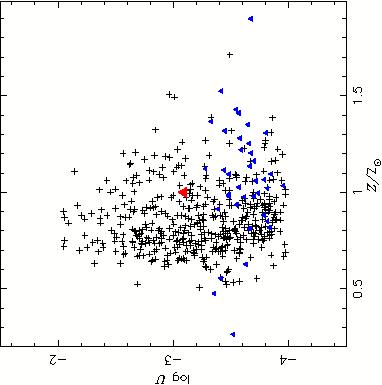}
\caption{As Fig.~\ref{figsl} but for logarithm of the ionzation
parameter ($\log U$) versus the metallicity ($Z/\rm Z_{\odot}$)
calculated by using the \citet{2020MNRAS.492.5675C} calibrations
(Eqs.~\ref{eq1} and \ref{eq2}).}
\label{figaa}
\end{figure}

According to Fig.~\ref{figsl}, LINERs have intermediate and low
[\ion{O}{iii}]/[\ion{O}{ii}] line ratio intensities, with the high 
values [$\rm (\log[\ion{O}{iii}]/[\ion{O}{ii}]) \: \ga \: 0.0$] only
observed in Seyfert~2.  Since the [\ion{O}{iii}]/[\ion{O}{ii}] has a strong
dependence on $U$, the above results indicate a tendency of LINERs
to present  lower $U$ values than the ones in Seyfert~2, as suggested by 
\cite{1983ApJ...264..105F}. As an additional test of this scenario, Fig.~\ref{figaa} presents  $\log U$ versus $Z/\rm Z_{\odot}$,
calculated by using the \citet{2020MNRAS.492.5675C} calibrations
(Eqs.~\ref{eq1} and \ref{eq2}), for the same sample as the one
in Fig.~\ref{figsl}. We can see that the UGC\,4805 and the LINERs 
occupy the region with lower $U$ values and the highest values
of this parameter are only observed in Seyfert~2s.

Finally, the  geometry of UGC\,4805 nucleus can provide
information about the ionization source. In view of this, 
we compare the ionization parameter 
derived from the AGN and pAGB photoionization models with 
the one estimated from the observational data. The average
value from the models is $<\log U>\sim -3.30$. To calculate
$U$ from observational data, first, we obtained the  $Q(\rm H)$
from the expression of \citet{2018MNRAS.479.3966H}
\begin{equation}
    \left(\frac{Q({\rm H)}}{ {\rm s^{-1}}}\right)=1.03\times10^{12}\left(\frac{L_{\rm H_{\alpha}}}{\rm s^{-1}}\right)
\end{equation}
and employing the luminosity value listed in Table~\ref{fluxos}.
This luminosity value is obtained from integrated flux of the
UGC\,4805 nucleus.
We found
$\log Q({\rm H)}=50.87$. The value $N_{\rm e}=100 \rm \: cm^{-3}$ is obtained
from [\ion{S}{ii}]$\lambda$6716/$\lambda6731$ line ratio intensity,  also
listed in Table~\ref{fluxos}. Applying the $Q(\rm H)$ and $N_{\rm e}$
values above to Eq.~\ref{elogu}, the innermost  radius value
$R_{0}$ to conciliate the theoretical and observational $U$ value 
is  about  50 pc, in order of the
radius assumed by \citet{2006A&A...456..953B}.
As can be noted in Fig.~\ref{bpt_diag}, the LINER emission extends to
until $\sim 2.5$ kpc, i.e.,  a high excitation level (or $U$) is maintained
from $\sim 50$ pc to kpc scales. Since $U\approx R^{-2}$,
the ionization source is probably spread along the $R$. Thus,  
this result  indicates that 
p-AGB is the preferable  ionization source rather than AGN. This assumption
is  supported by the result obtained previously from the WHAN diagram \citep{2011MNRAS.413.1687C}.


\section{Conclusion}
\label{conc}

We used optical emission-line fluxes  taken from the SDSS-IV MaNGA survey to determine the oxygen abundance (metallicity) of  the LINER nucleus of the UGC\,4805 galaxy. The oxygen abundance was derived through the extrapolation of the radial abundance gradient for the central part of the disk by using strong-line calibrations for AGNs and  photoionization model grids assuming as ionizing sources gas accretion into a black hole, representing an AGN and p-AGB stars. We found that all the O/H abundance estimations agree with each other.
The results from these methods indicate  that the 
UGC\,4805 nucleus  has an oxygen abundance in the range of  $1.0 \: \la \: (Z/Z_{\odot}) \: \la 1.35$, i.e., solar or slightly over-solar metallicity.

We calculated that the UGC\,4805 nucleus and other LINERs present metallicity and ionization parameter sensitive emission-line ratios similar to those observed in
confirmed Seyfert 2 nuclei,l although exhibiting a slightly different trend. Even though LINERs present low ionization parameter values ($\log U \: \la \: -3.2$), Seyfert~2 nuclei also present low values of the ionization parameter. 
Although both AGN and p-AGB models (with $T_{\rm eff}$= 100 and 190 kK) are able to reproduce the observational data, the results from the WHAN diagram combined with the fact that the high  excitation level of the gas has to be maintained at kpc scales, suggest
that the main ionizing source of the UGC\,4805 nucleus probably has a stellar origin rather than an AGN.

\section*{Acknowledgements}
 ACK thanks to CNPq. CBO is grateful to the FAPESP for the support under grant 2019/11934-0, and to the CAPES. IAZ acknowledges support by the grant for young scientist's research laboratories of the National Academy of Sciences of Ukraine. AHJ thanks to CONICYT, Programa de Astronom\'ia,
Fondo ALMA-CONICYT 2017, C\'odigo de proyecto 31170038.



\section{Data Availability}

The data underlying this article will be shared on reasonable request
to the corresponding author.

\bibliographystyle{mnras}
\bibliography{krabbe} 



\bsp	
\label{lastpage}
\end{document}